\documentclass[aps,prd,reprint,preprintnumbers,nofootinbib]{revtex4-1}

\usepackage[utf8]{inputenc}
\usepackage{amsmath}
\usepackage{amsfonts}
\usepackage{amssymb}
\usepackage{color}
\usepackage{graphicx}
\usepackage{bbold}
\usepackage{mathtools}

\usepackage[colorlinks]{hyperref}
\hypersetup{allcolors=[rgb]{0,0,1}}

\newcommand{\ket}[1]{|#1\rangle}
\newcommand{\eps}{\epsilon}
\newcommand{\orcid}[1]{\href{https://orcid.org/#1}{#1}}

\newcommand{\dmsqee}{\Delta m^2_{ee}}
\renewcommand{\c}{c_{(\tilde{\theta}_{13}-\theta_{13})}}
\newcommand{\s}{s_{(\tilde{\theta}_{13}-\theta_{13})}}
\newcommand{\Dl}[2]{\Delta\lambda_{#1#2}}

\begin{document}

\title{Rotations Versus Perturbative Expansions for Calculating Neutrino Oscillation Probabilities in Matter}

\author{Peter B.~Denton}
\email{peterbd1@gmail.com}
\thanks{\orcid{0000-0002-5209-872X}}
\affiliation{Niels Bohr International Academy, University of Copenhagen, The Niels Bohr Institute, Blegdamsvej 17, DK-2100, Copenhagen, Denmark}

\author{Stephen J.~Parke}
\email{parke@fnal.gov}
\thanks{\orcid{0000-0003-2028-6782}}
\affiliation{Theoretical Physics Department, Fermi National Accelerator Laboratory, P.~O.~Box 500, Batavia, IL 60510, USA}

\author{Xining Zhang}
\email{xining@uchicago.edu}
\thanks{\orcid{0000-0001-8959-8405}}
\affiliation{Enrico Fermi Institute and Department of Physics, University of Chicago, Chicago, Illinois 60637, USA}

\preprint{FERMILAB-PUB-18-213-T, IFT-UAM/CSIC-18-54}

\date{June 29, 2018}

\begin{abstract}
We further develop a simple and compact technique for calculating the three flavor neutrino oscillation probabilities in uniform matter density. By performing additional rotations instead of implementing a perturbative expansion we significantly decrease the scale of the perturbing Hamiltonian and therefore improve the accuracy of zeroth order. We explore the relationship between implementing additional rotations and that of performing a perturbative expansion. Based on our analysis, independent of the size of the matter potential, we find that the first order perturbation expansion can be replaced by two additional rotations and a second order perturbative expansion can be replaced by one more rotation. Numerical tests have been applied and all the exceptional features of our analysis have been verified.
\end{abstract}

\maketitle

\section{Introduction}
After Wolfenstein showed that neutrino oscillations are altered in matter, \cite{Wolfenstein:1977ue} exact analytic solutions for three flavors were calculated under the assumption of uniform matter density \cite{Barger:1980tf,Zaglauer:1988gz}. However, the exact solutions are too complex to understand in practice leading to an interest in alternative approaches including perturbative expansions. One possible expansion parameter is $\sin\theta_{13}$ \cite{Cervera:2000kp,Minakata:2009sr,Asano:2011nj}, but we now know that $\sin\theta_{13}=0.13$ \cite{An:2016bvr,Ahn:2012nd} is not as small as was anticipated making these expansions very lengthy in order to reach acceptable levels of precision. Moreover, when expanding around $\sin\theta_{13}=0$, two of the eigenvalues cross at an energy around $E\sim 10$ GeV for Earth density, thus a perturbative expansion will not converge near the atmospheric resonance. The only other available choice of an expansion parameter
is $\Delta m^2_\odot/\Delta m^2_\oplus\simeq 0.03$, for arbitrary size of the matter potential, but this choice also
has a similar issue of crossing eigenvalues at the solar resonance, near $E\sim 140$ MeV
for Earth density, and thus such a perturbative expansion will not converge near the
solar resonance \cite{Cervera:2000kp,Arafune:1996bt,Freund:2001pn,Akhmedov:2004ny,Minakata:2015gra}.
For a perturbative expansion to be effective for all values of the matter potential,
one has to deal with these two level crossings in a non-perturbative manner first. This
is achieved by performing rotations in the (1-3) and (1-2) sectors so that diagonal
values of the Hamiltonian do not cross for any value of the matter potential. This
was first performed in \cite{Blennow:2013rca,Denton:2016wmg}. When performing the (1-3) rotation, it is very
natural to absorb part of the sub-leading terms into the zeroth order by using
\begin{align}
\Delta m^2_{ee}&\equiv \cos^2\theta_{12}\Delta m^2_{31}+\sin^2\theta_{12}\Delta m^2_{32}\notag\\
&=\Delta m^2_{31}-\sin^2\theta_{12}\Delta m^2_{21},
\end{align}
instead of $\Delta m^2_{31}$, see \cite{Minakata:2015gra}. This is the atmospheric $\Delta m^2$ measured in a $\nu_e$ disappearance experiment \cite{Nunokawa:2005nx,Parke:2016joa}.

After both the (1-3) and (1-2) rotations, given in \cite{Denton:2016wmg}, the expansion parameter
for the perturbing Hamiltonian is
\begin{gather}
\epsilon'\equiv\epsilon\sin(\tilde{\theta}_{13}-\theta_{13})\sin\theta_{12}\cos\theta_{12},\notag\\
\text{where}\notag\\
\epsilon\equiv\Delta m^2_{21}/\Delta m^2_{ee}\simeq 0.03,
\end{gather}
and $\tilde{\theta}_{13}$ is the value of the mixing angle, $\theta_{13}$, in matter.  Thus the magnitude of the expansion parameter
is never larger than 0.015 and vanishes in vacuum. After these two two-flavor
rotations, the perturbative expansion is well behaved for all values of the matter
potential and zeroth, first and second order perturbative results are all given in \cite{Denton:2016wmg}.

In this paper, we further develop the method in \cite{Denton:2016wmg}. We will perform additional rotations such that the scale of the perturbing Hamiltonian will be significantly decreased. Accordingly, the accuracy of the zeroth order Hamiltonian will be improved. The advantages of the former works are inherited, i.e.~the additional rotations defined here continue to be valid for all channels, any terrestrial or solar matter potential, and the new rotation matrices return to the identity in vacuum. It is reasonable that if the perturbing Hamiltonian is small enough, in another word the zeroth order Hamiltonian is sufficiently accurate, the zeroth order expressions are already a good enough approximation such that perturbation theory is no longer required. We prove that two additional rotations can take the place of a first order perturbation theory and a second order perturbation theory can be replaced by three additional rotations.  In principle, performing additional rotations can be chosen to be equivalent to any order of the perturbation expansion, although unnecessary for the expected precision of any future oscillation experiment.

The structure of this paper is listed following. In section \ref{sec:DMP}, we briefly review the method developed in \cite{Denton:2016wmg}. The general principles to perform additional rotations are enumerated. Section \ref{sec:additional rotations} includes the main results of this paper. We provide details to determine sequence of the addition rotations and values of the rotation angles; the zeroth order eigenvalues and eigenstates after the rotations. We also compare the additional rotations with the perturbation theories and prove the equivalence order by order in this section. In section \ref{sec:corrections to angles} we calculate the corrected mixing angles and CP phase in matter. Finally the conclusion is in section \ref{sec:conclusions}. All other remarks and supplementary materials we believe necessary can be found in the Appendices. 

\section{Zeroth order approximation of neutrino propagation in matter}
\label{sec:DMP}
This section reviews \cite{Denton:2016wmg} through zeroth order. The Schr\"odinger equation governing neutrino evolution in matter is
\begin{equation}
i\frac{\partial}{\partial x}\ket{\nu}=H\ket{\nu}. \label{eq:sch}
\end{equation}
In the flavor basis $\ket{\nu}=(\nu_e,\nu_\mu,\nu_\tau)^T$, the Hamiltonian is
\begin{multline}
H=\frac{1}{2E}\left[U_\text{PMNS}\text{diag}(0,\Delta m^2_{21},\Delta m^2_{31})U^\dagger_\text{PMNS}\right.\\
\left.+\text{diag}(a(x),0,0)\right].
\end{multline}
The lepton mixing matrix in vacuum $U_\text{PMNS}$ \cite{Maki:1962mu,Pontecorvo:1967fh} is defined by the product of a sequence of rotation matrices in 23, 13, and 12 plane, i.e.~$U_\text{PMNS}\equiv U_{23}(\theta_{23},\delta)U_{13}(\theta_{13})U_{12}(\theta_{12})$, in which the $U_{23}$ rotation is a complex rotation with a complex phase $\delta$, the PDG form of $U_\text{PMNS}$ can be obtained from our $U_\text{PMNS}$ by multiplying the 3rd row by $e^{i\delta}$ and the 3rd column by $e^{-i\delta}$. The matter potential is assumed to be a constant $a(x)=a\equiv 2\sqrt{2}G_FN_eE$.

Eq.~\ref{eq:sch} still holds if both sides are multiplied by some constant unitary matrix $U^\dagger$ simultaneously, and since $UU^\dagger$ is the identity matrix we are free to insert it between $H$ and $\ket{\nu}$ on the right hand side. The transformed neutrino basis is 
\begin{equation}
\ket{\check{\nu}}=U^\dagger\ket{\nu}, \label{eq:zerothstates}
\end{equation}
and in this basis the Hamiltonian is 
\begin{equation}
\check{H}=U^\dagger HU,
\end{equation}
where some appropriate unitary matrix $U$ such that the transformed Hamiltonian $\check{H}$ satisfies the following three properties:
\begin{itemize}
\item The diagonal elements are good approximations to the exact eigenvalues.
\item The off-diagonal elements are small.
\item $\check{H}$ is identical to $\text{diag}(0,\Delta m^2_{21},\Delta m^2_{31})$ in vacuum.
\end{itemize}
Thus the diagonal elements of $\check{H}$ are zeroth order approximations to the eigenvalues. If the unitary matrix $U$ can also be written as the product of a sequence of rotations matrices as $U_\text{PMNS}$, i.e.~$U=U_{23}(\tilde{\theta}_{23},\tilde{\delta})U_{13}(\tilde{\theta}_{13})U_{12}(\tilde{\theta}_{12})$, the angles $\{\tilde{\theta}_{23},\tilde{\theta}_{13},\tilde{\theta}_{12}\}$ are zeroth order approximations to the three mixing angles in matter, and $\tilde{\delta}$ is the zeroth order CP phase in matter. The calculation process of these zeroth order values are summarized in Appendix \ref{Appendixzerothorder}, more details can be found in \cite{Denton:2016wmg}. Here we just cite the results. 

The zeroth order approximation of the (2-3) mixing angle and the CP phase in matter are
\begin{align}
\tilde{\theta}_{23}&=\theta_{23}, \label{eq:theta}\\
\tilde{\delta}&=\delta. \label{eq:delta}
\end{align}
The (1-3) mixing angle in matter is determined by
\begin{equation}
\tan 2\tilde{\theta}_{13}=\frac{s_{2\theta_{13}}\Delta m^2_{ee}}{c_{2\theta_{13}}\Delta m^2_{ee}-a}, \quad \tilde{\theta}_{13}\in [0,\pi/2].
\end{equation}
The (1-2) mixing angle in matter is determined by
\begin{equation}
\tan 2\tilde{\theta}_{12}=\frac{\epsilon c_{(\tilde{\theta}_{13}-\theta_{13})}s_{2\theta_{12}}\Delta m^2_{ee}}{\lambda_0-\lambda_-}, \label{eq:sin2theta12} \quad\tilde{\theta}_{12}\in [0,\pi/2],
\end{equation}
where
\begin{multline}
\lambda_0-\lambda_-=\epsilon c_{2\theta_{12}}\Delta m^2_{ee}-\frac{1}{2}\left[a+\Delta m^2_{ee}\right.\\
\left.-\text{sign}(\Delta m^2_{ee})\sqrt{(c_{2\theta_{13}}\Delta m^2_{ee}-a)^2+(s_{2\theta_{13}}\Delta m^2_{ee})^2}\right].
\end{multline}

\begin{figure*}[t]
\centering
\includegraphics[scale=0.5]{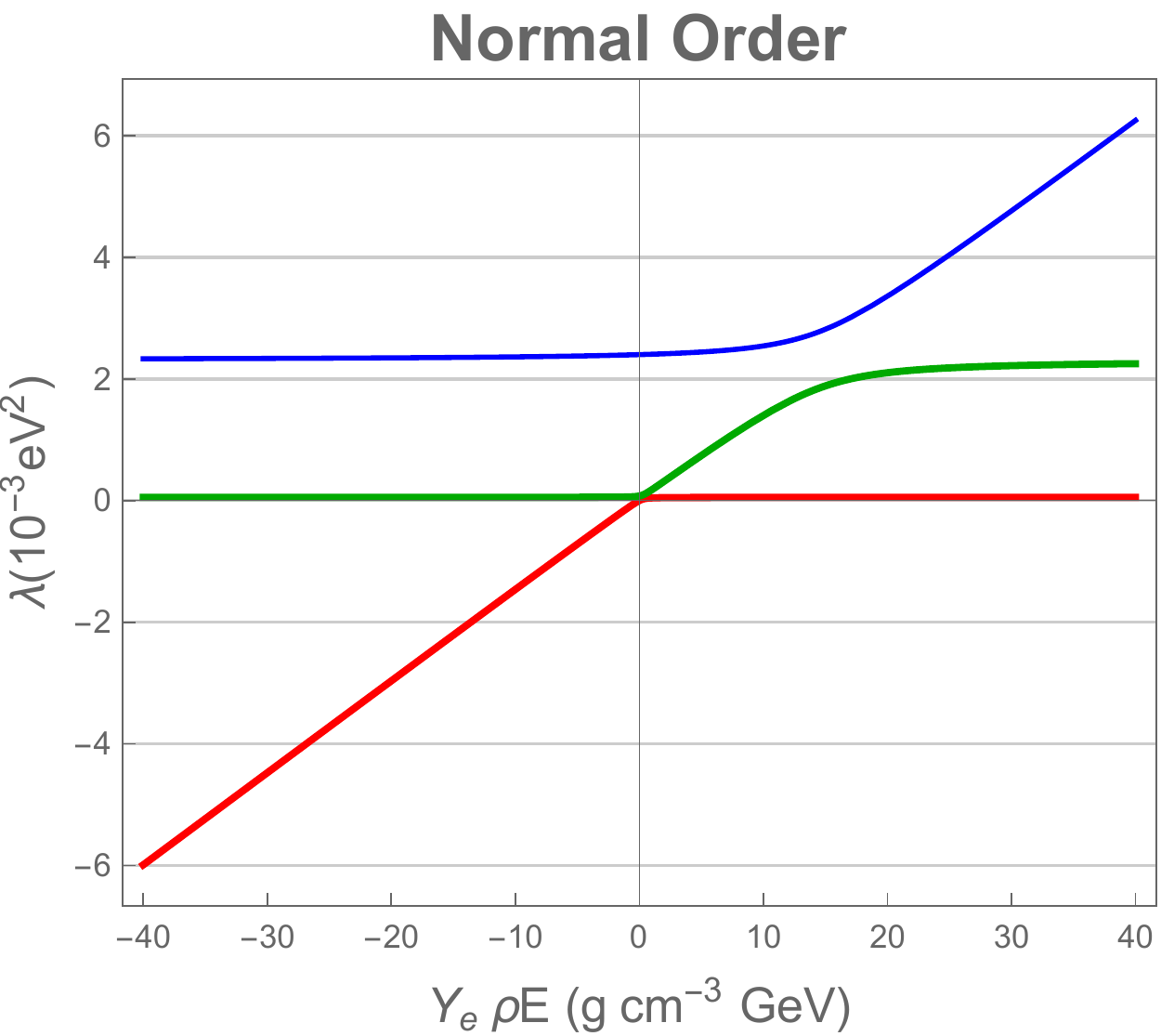}
\includegraphics[scale=0.5]{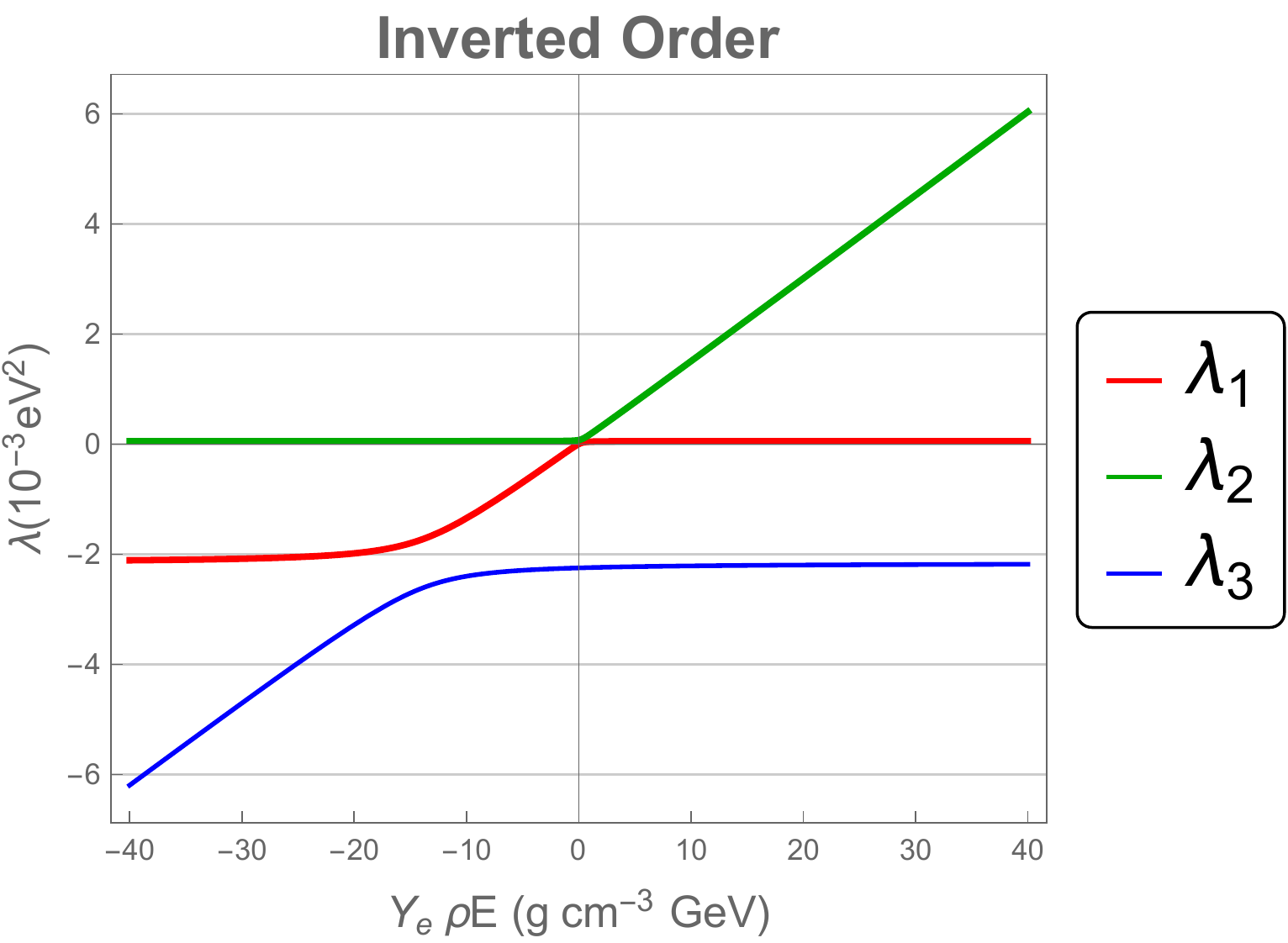}
\includegraphics[scale=0.5]{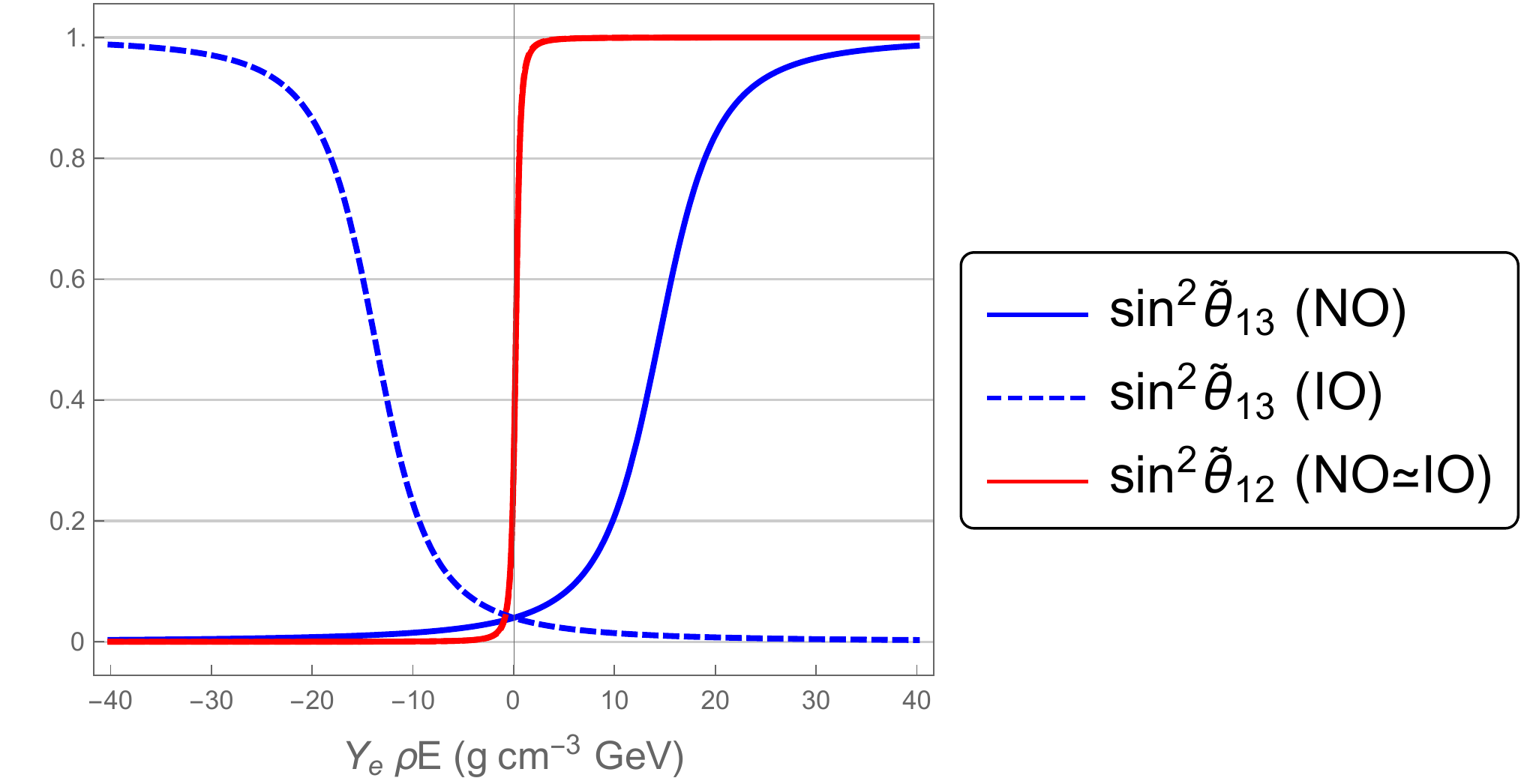}
\caption{The upper two figures show the eigenvalues to zeroth order in matter as functions of the matter potential. The upper-left plot is for normal mass ordering and the upper-right plot is for inverted mass order. The lower plot shows the mixing angles $\sin^2\tilde{\theta}_{12}$, $\sin^2\tilde{\theta}_{13}$ to zeroth order in matter, and the solid (dashed) curves are for normal (inverted) mass ordering. For $\sin^2\tilde{\theta}_{12}$, the curves of both mass orders overlap but are not identical. }
\label{fig:zerothordervalues} 
\end{figure*}

Finally $\check{H}$ can be expressed as 
\begin{equation}
\check{H}=\underbrace{\frac{1}{2E}\left(\begin{array}{ccc}
\lambda_1& & \\
&\lambda_2 & \\
& &\lambda_3 \\
\end{array}
\right)}_{\check{H}_0}+\underbrace{\epsilon^\prime \frac{\Delta m^2_{ee}}{2E}\left(\begin{array}{ccc}
& &-\tilde{s}_{12} \\
& & \tilde{c}_{12} \\
-\tilde{s}_{12} & \tilde{c}_{12} & \\
\end{array}
\right)}_{\check{H}_1}, \label{eq:Hbar}
\end{equation}
where 
\begin{equation}
\epsilon^\prime\equiv \epsilon s_{(\tilde{\theta}_{13}-\theta_{13})}s_{12}c_{12}, \qquad |\epsilon^\prime|<0.015,
\end{equation}
and $\tilde{s}_{ij}$, $\tilde{c}_{ij}$ represent $\sin\tilde{\theta}_{ij}$, $\cos\tilde{\theta}_{ij}$, respectively. $\lambda_i$ are the diagonal elements of the Hamiltonian after the $U_{12}(\tilde{\theta}_{12})$ rotation, they are also the zeroth order approximations to the eigenvalues in matter, their values can be found in Eq.~\ref{eq:lambdapm0}, Appendix~\ref{Appendixzerothorder}. $\check{H}_0$ is the zeroth order Hamiltonian and $\check{H}_1$ is the perturbing term and $\eps'=0$ in vacuum. 
The numerical values of the zeroth order eigenvalues and mixing angles are plotted in Fig.~\ref{fig:zerothordervalues}.

\section{Additional rotations}
\label{sec:additional rotations}
Ref.~\cite{Denton:2016wmg} presented a general principle to enhance zeroth order accuracy by performing a rotation diagonalizing the sector with leading order off-diagonal entries. Therefore at the point of Eq.~\ref{eq:Hbar}, we can perform additional rotations to further improve the zeroth order. This idea is initialized in \cite{Parke:2018brr}. 

Since $\check{H}_0$ is diagonal, to determine the leading order off-diagonal entries we just need to study $\check{H}_1$, more specifically, we compare $\tilde{s}_{12}$ and $\tilde{c}_{12}$. The red curve in the lower panel of Fig.~\ref{fig:zerothordervalues} shows how $\tilde{s}_{12}$ depends on the matter potential. For large matter effect we have that $|\tilde{s}_{12}|\gg|\tilde{c}_{12}|$ when $Y_e\rho E\gg 0$ and $|\tilde{s}_{12}|\ll|\tilde{c}_{12}|$ when $Y_e\rho E\ll 0$. However, when the matter potential is weak we must be more careful since $\tilde{s}_{12}$ and $\tilde{c}_{12}$ are close in this case. We find that $|\tilde{s}_{12}|=|\tilde{c}_{12}|=1/\sqrt{2}$ when $Y_e\rho E\simeq 0.2$ g cm$^{-3}$ GeV where we have taken $s^2_{12}\simeq 0.3$ \cite{GonzalezGarcia:2012sz}.
This critical point is applicable to both normal and inverted mass orderings. When the matter effect is weak, $\tilde{\theta}_{13}\simeq \theta_{13}$ so $s_{(\tilde{\theta}_{13}-\theta_{13})}\simeq 0$ so $\eps'\simeq0$. Then $\check{H}_1$ will be a higher order perturbation which is small. Therefore the convenience of additional rotations depends only on the sign of $Y_e\rho E$, i.e.~the case of neutrinos or anti-neutrinos. 

In general, the diagonalizing angle is given by the simple expression
\begin{equation}
\tan2\theta=\frac{2\lambda_x}{\lambda_b-\lambda_a},
\end{equation}
where $\lambda_x$ is the off-diagonal part and $\lambda_a$ ($\lambda_b$) is the first (second) row diagonal element in the 2$\times$2 sub-matrix to be diagonalized.
The two new eigenvalues are
\begin{align}
\lambda_\sigma&=c_\theta^2\lambda_a+s_\theta^2\lambda_b-2s_\theta c_\theta\lambda_x,\notag\\
\lambda_\rho&=s_\theta^2\lambda_a+c_\theta^2\lambda_b+2s_\theta c_\theta\lambda_x, \label{eq:generaldiagonalize}
\end{align}
and the third eigenvalue remains the same.
For $|\lambda_x|\ll|\Delta\lambda_{ba}|$, $\theta$ is small, so we can expand this to get
\begin{align}
\lambda_\sigma&\simeq\lambda_a-\frac{\lambda_x^2}{\Delta\lambda_{ba}}\left\{1+\mathcal{O}\left[\left(\frac{\lambda_x}{\Delta\lambda_{ba}}\right)^2\right]\right\},\notag\\
\lambda_\rho&\simeq\lambda_b+\frac{\lambda_x^2}{\Delta\lambda_{ba}}\left\{1+\mathcal{O}\left[\left(\frac{\lambda_x}{\Delta\lambda_{ba}}\right)^2\right]\right\},\label{eq:generaldiagonalizeapproximate}
\end{align}
where $\Delta\lambda_{ij}=\lambda_i-\lambda_j$. More details can be found in Appendix A.1 in \cite{Denton:2016wmg}.

It is clear from Eq.~\ref{eq:generaldiagonalize} that performing a rotation leaves the trace (sum of eigenvalues) unchanged, and therefore, the trace remains unchanged through first order in the smallness parameter as shown in Eq.~\ref{eq:generaldiagonalizeapproximate}.

\subsection{Neutrino case}
In the case of neutrinos, $Y_e\rho E$ is positive, which means $|\tilde{s}_{12}|\gtrsim|\tilde{c}_{12}|$. Thus we will rotate in (1-3) sector first. We will then show that after the first rotation in (1-3) sector, the second and third rotations will be in (2-3) and (1-2) sectors, respectively.
\subsubsection{\texorpdfstring{$U_{13}$}{U13} rotation}
Define $\alpha_{13}$ to be the next rotation angle. The Hamiltonian after the $U_{13}(\alpha_{13})$ rotation is defined as
\begin{equation}
\check{H}^\prime\equiv U^\dagger_{13}(\alpha_{13})\check{H}U_{13}(\alpha_{13}). 
\end{equation}
Detailed formula of $\check{H}^\prime$ can be found in Eqs.~\ref{eq:checkH0p},~\ref{eq:checkH1p}, Appendix \ref{Appendixadditionalrotations}. The rotation angle diagonalizing the (1-3) sector is:
\begin{equation}
\alpha_{13}=-\frac{1}{2}\arctan\frac{2\epsilon^\prime\Delta m_{ee}^2\tilde{s}_{12}}{\Delta\lambda_{31}}\simeq-\frac{\epsilon^\prime\Delta m_{ee}^2\tilde{s}_{12}}{\Delta\lambda_{31}}+\mathcal{O}(\epsilon^{\prime\,3}). \label{eq:alpha13}
\end{equation}
Since $\Delta\lambda_{31}\gtrsim\Delta m^2_{ee}$, $\alpha_{13}$ is at least first order in $\eps'$. The diagonal elements, $\lambda^\prime_i$, are the new zeroth order eigenvalues. They are
\begin{align}
\lambda^\prime_1&=c^2_{\alpha_{13}}\lambda_1+s^2_{\alpha_{13}} \lambda_3+2s_{\alpha_{13}}c_{\alpha_{13}}\tilde{s}_{12} \epsilon^\prime \Delta m^2_{ee}\notag\\
&\simeq \lambda_1-(\epsilon^\prime\Delta m^2_{ee})^2\frac{\tilde{s}^2_{12}}{\Delta\lambda_{31}}+\mathcal{O}(\epsilon^{\prime\,4}), \notag\\
\lambda^\prime_2&=\lambda_2, \notag\\
\lambda^\prime_3&=s^2_{\alpha_{13}}\lambda_1+c^2_{\alpha_{13}}\lambda_3-2s_{\alpha_{13}}c_{\alpha_{13}}\tilde{s}_{12} \epsilon^\prime \Delta m^2_{ee}\notag\\
&\simeq \lambda_3+(\epsilon^\prime\Delta m^2_{ee})^2\frac{\tilde{s}^2_{12}}{\Delta\lambda_{31}}+\mathcal{O}(\epsilon^{\prime\,4}). \label{eq:lambdaP}
\end{align}

It is remarkable to notice that the additional rotation $U_{13}(\alpha_{13})$ does not make first order (in $\epsilon^\prime$) corrections to the eigenvalues. This conclusion agrees with a first order perturbation theory. It is known that in perturbation theories first order corrections to eigenvalues are just the diagonal elements of the perturbing Hamiltonian. Since all diagonal entries of $\check{H}_1$ vanish, the first order corrections are zero. This equivalence indicates a close relation between the additional rotations and perturbation theory discussed in further detail in Sec.~\ref{comparewithperturbation}.

\subsubsection{\texorpdfstring{$U_{23}$}{U23} rotation}
Since $\alpha_{13}$ is small, the leading order off-diagonal entries in the Hamiltonian are proportional to $\tilde{c}_{12}c_{\alpha_{13}}$ so the next rotation should diagonalize the (2-3) sector with a new angle $\alpha_{23}$. The rotated Hamiltonian is
\begin{equation}
\check{H}^{\prime\prime}\equiv U^\dagger_{23}(\alpha_{23})U^\dagger_{13}(\alpha_{13})\check{H} U_{13}(\alpha_{13})U_{23}(\alpha_{23}), 
\end{equation}
detailed formula can be found in Eqs.~\ref{eq:checkH0pp},~\ref{eq:checkH1pp} in Appendix~\ref{Appendixadditionalrotations}. The rotation angle is
\begin{equation}
\alpha_{23}=\frac{1}{2}\arctan\frac{2\epsilon^\prime\Delta m_{ee}^2c_{\alpha_{13}}\tilde{c}_{12}}{\Delta\lambda^\prime_{32}}\simeq\frac{\epsilon^\prime\Delta m_{ee}^2\tilde{c}_{12}}{\Delta\lambda_{32}}+\mathcal{O}(\epsilon^{\prime\,3}). \label{eq:alpha23}
\end{equation}
As with $\alpha_{31}$, $\alpha_{32}$ is also at least first order in $\eps'$ since $\Delta\lambda_{32}\gtrsim\Delta m^2_{ee}$. The new eigenvalues are
\begin{align}
\lambda^{\prime\prime}_1&=\lambda^\prime_1 \simeq \lambda_1-(\epsilon^\prime\Delta m^2_{ee})^2\frac{\tilde{s}^2_{12}}{\Delta\lambda_{31}}+\mathcal{O}(\epsilon^{\prime\,4}), \notag\\
\lambda^{\prime\prime}_2&=c^2_{\alpha_{23}}\lambda^\prime_2+s^2_{\alpha_{23}} \lambda^\prime_3-2s_{\alpha_{23}}c_{\alpha_{23}}c_{\alpha_{13}}\tilde{c}_{12} \epsilon^\prime \Delta m^2_{ee}\notag\\
&\simeq\lambda_2-(\epsilon^\prime\Delta m^2_{ee})^2\frac{\tilde{c}^2_{12}}{\Delta\lambda_{32}}+\mathcal{O}(\epsilon^{\prime\,4}),\notag\\
\lambda^{\prime\prime}_3&=s^2_{\alpha_{23}}\lambda^\prime_2+c^2_{\alpha_{23}} \lambda^\prime_3+2s_{\alpha_{23}}c_{\alpha_{23}}c_{\alpha_{13}}\tilde{c}_{12} \epsilon^\prime \Delta m^2_{ee}\notag\\
&\simeq \lambda_3+(\epsilon^\prime\Delta m^2_{ee})^2\left(\frac{\tilde{s}^2_{12}}{\Delta\lambda_{31}}+\frac{\tilde{c}^2_{12}}{\Delta\lambda_{32}}\right)+\mathcal{O}(\epsilon^{\prime\,4}). \label{eq:lambdaPP}
\end{align}

\subsubsection{\texorpdfstring{$U_{12}$}{U12} rotation}
Again $\alpha_{23}$ is small so it is evident that after the $U_{23}(\alpha_{23})$ rotation the leading order off-diagonal entries, which proportional to $s_{\alpha_{13}}c_{\alpha_{23}}\tilde{c}_{12}$, are in the (1-2) sector, and an additional rotation $U_{12}(\alpha_{12})$ can diagonalize it. The final rotated Hamiltonian is 
\begin{multline}
\check{H}^{\prime\prime\prime}\equiv U^\dagger_{12}(\alpha_{12})U^\dagger_{23}(\alpha_{23})U^\dagger_{13}(\alpha_{13})\notag\\
\times\check{H}U_{13}(\alpha_{13})U_{23}(\alpha_{23}) U_{12}(\alpha_{12}). 
\end{multline}
Again details of $\check{H}^{\prime\prime\prime}$ can be found in Eqs.~\ref{eq:checkH0ppp},~\ref{eq:checkH1ppp} in Appendix~\ref{Appendixadditionalrotations}. It can be solved that
\begin{align}
\alpha_{12}&=-\frac{1}{2}\arctan\frac{2\epsilon^\prime\Delta m_{ee}^2 c_{\alpha_{23}}s_{\alpha_{13}}\tilde{c}_{12}}{\Delta\lambda^{\prime\prime}_{21}}\notag\\
&\simeq\frac{(\epsilon^\prime\Delta m_{ee}^2)^2\tilde{s}_{12} \tilde{c}_{12}}{\Delta\lambda_{21}\Delta\lambda_{31}}+\mathcal{O}(\epsilon^{\prime\,4}). \label{eq:alpha12}
\end{align}
The zeroth order eigenvalues, after the (1-2) rotation are 
\begin{align}
\lambda^{\prime\prime\prime}_1&=c^2_{\alpha_{12}}\lambda^{\prime\prime}_1+s^2_{\alpha_{12}} \lambda^{\prime\prime}_2+2s_{\alpha_{12}}c_{\alpha_{12}}c_{\alpha_{23}}s_{\alpha_{13}}\tilde{c}_{12} \epsilon^\prime \Delta m^2_{ee}\notag\\
&\simeq \lambda_1-(\epsilon^\prime\Delta m^2_{ee})^2\frac{\tilde{s}^2_{12}}{\Delta\lambda_{31}}+\mathcal{O}(\epsilon^{\prime\,4}),\notag\\
\lambda^{\prime\prime\prime}_2&=s^2_{\alpha_{12}}\lambda^{\prime\prime}_1+c^2_{\alpha_{12}} \lambda^{\prime\prime}_2-2s_{\alpha_{12}}c_{\alpha_{12}}c_{\alpha_{23}}s_{\alpha_{13}}\tilde{c}_{12} \epsilon^\prime \Delta m^2_{ee}\notag\\
&\simeq \lambda_2-(\epsilon^\prime\Delta m^2_{ee})^2\frac{\tilde{c}^2_{12}}{\Delta\lambda_{32}}+\mathcal{O}(\epsilon^{\prime\,4}),\notag\\
\lambda^{\prime\prime\prime}_3&= \lambda^{\prime\prime}_3\simeq \lambda_3+(\epsilon^\prime\Delta m^2_{ee})^2\left(\frac{\tilde{s}^2_{12}}{\Delta\lambda_{31}}+\frac{\tilde{c}^2_{12}}{\Delta\lambda_{32}}\right)+\mathcal{O}(\epsilon^{\prime\,4}).\label{eq:lambdaPPP}
\end{align}
It is noteworthy that $\lambda^{\prime\prime\prime}_i$ and $\lambda^{\prime\prime}_i$ are identical to at least second order. To understand this observation, we need to study the perturbative Hamiltonians after each rotation. It is known that in a perturbative expansion, leading order corrections to the eigenvalues are the diagonal elements of the perturbative Hamiltonian. In Appendix~\ref{Appendixadditionalrotations}, we shall demonstrate that after the first two additional rotations, the perturbative Hamiltonian whose diagonal entries are all zero, is in second order; thus errors of $\lambda^{\prime\prime}_i$ are already controlled to fourth order. After the third rotation $U_{12}(\alpha_{12})$, the perturbative Hamiltonian (still with vanishing diagonal entries) is in third order; thus errors of $\lambda^{\prime\prime\prime}_i$ are further diminished to sixth order. Therefore, it is not unexpected that $\lambda^{\prime\prime}_i$ and $\lambda^{\prime\prime\prime}_i$ are identical to second order.

Terms of order $\epsilon^{\prime\,3}$ are no larger than $3\times10^{-6}$. In principle, we can continue performing rotations to control the off-diagonal entries to any precision. Considering the precision of the experimental uncertainties $\sim1\%$ \cite{Acciarri:2015uup,Patterson:2012zs,Abe:2011ks,Abe:2015zbg,Kelly:2018kmb}, stopping at $U_{12}(\alpha_{12})$ is more than enough. Later we will show that it is equal to second order (in $\epsilon^\prime$) perturbation theory when considering eigenstates. 

\subsection{Anti-neutrino case}
In the case where $Y_e\rho E\lesssim 0$, $|\tilde{s}_{12}|\lesssim|\tilde{c}_{12}|$ in $\check H_1$ of Eq.~\ref{eq:Hbar}, so we will rotate (2-3) sector before (1-3), and the third additional rotation will still be in (1-2) sector as for neutrinos. The calculation process will be quite similar to the first case. The results for this case are listed below. The (2-3) rotation angle is
\begin{equation}
\bar{\alpha}_{23}=\frac{1}{2}\arctan\frac{2\epsilon^\prime\Delta m_{ee}^2\tilde{c}_{12} }{\Delta\lambda_{32}}\simeq\frac{\epsilon^\prime\Delta m_{ee}^2\tilde{c}_{12}}{\Delta\lambda_{32}}+\mathcal{O}(\epsilon^{\prime\,3}). 
\end{equation}
Compared with Eq.~\ref{eq:alpha23}, it is evident that $\alpha_{23}\simeq\bar{\alpha}_{23}$ to first order. After the (2-3) rotation, the zeroth order eigenvalues are
\begin{align}
\bar{\lambda}^\prime_1&= \lambda_1, \notag\\
\bar{\lambda}^\prime_2&=c^2_{\bar{\alpha}_{23}}\lambda_2+s^2_{\bar{\alpha}_{23}}\lambda_3-2s_{\bar{\alpha}_{23}}c_{\bar{\alpha}_{23}}\tilde{c}_{12} \epsilon^\prime \Delta m^2_{ee}\notag\\
&\simeq \lambda_2-(\epsilon^\prime\Delta m^2_{ee})^2\frac{\tilde{c}^2_{12}}{\Delta\lambda_{32}}+\mathcal{O}(\epsilon^{\prime\,4}), \notag\\
\bar{\lambda}^\prime_3&=s^2_{\bar{\alpha}_{23}}\lambda_2+c^2_{\bar{\alpha}_{23}}\lambda_3+2s_{\bar{\alpha}_{23}}c_{\bar{\alpha}_{23}}\tilde{c}_{12} \epsilon^\prime \Delta m^2_{ee}\notag\\
&\simeq \lambda_3+(\epsilon^\prime\Delta m^2_{ee})^2\frac{\tilde{c}^2_{12}}{\Delta\lambda_{32}}+\mathcal{O}(\epsilon^{\prime\,4}). 
\end{align}
Before performing the next additional rotation in (1-3) sector, there are some comments on the above $U_{23}$ rotation. In some former works, e.g.~\cite{Blennow:2013rca}, a similar approach was followed with a rotation in the (2-3) sector as above, although there the rotation was used for both neutrinos and anti-neutrinos. In additional, later in this paper (section \ref{comparewithperturbation} and Fig.~\ref{fig:Schematic}), we shall demonstrate that one additional rotation does not improve the accuracy of the approximated eigenstates. More specifically, if $\ket{\check{\nu}}^m$ is the exact eigenstates in matter, errors of the initial zeroth order eigenstates are estimated as $\ket{\check{\nu}}^m-\ket{\check{\nu}}\simeq\mathcal{O(\epsilon^\prime)}$. After the $U_{23}$ rotation, the eigenstates are corrected to be $U^\dagger_{23}\ket{\check{\nu}}$, which still have first order errors, i.e. $\ket{\check{\nu}}^m-U^\dagger_{23}\ket{\check{\nu}}\simeq\mathcal{O(\epsilon^\prime)}$ still holds. This indicates that to achieve better accuracy, we must perform an additional rotation.

The following (1-3) rotation angle is
\begin{align}
\bar{\alpha}_{13}&=-\frac{1}{2}\arctan\frac{2\epsilon^\prime\Delta m_{ee}^2c_{\bar{\alpha}_{23}}\tilde{s}_{12}}{\Delta\bar{\lambda}^\prime_{31}}\notag\\
&\simeq-\frac{\epsilon^\prime\Delta m_{ee}^2\tilde{s}_{12}}{\Delta\lambda_{31}}+\mathcal{O}(\epsilon^{\prime\,3}). 
\end{align}
Again, compared with Eq.~\ref{eq:alpha13}, $\alpha_{13}\simeq\bar{\alpha}_{13}$ to first order. After the (1-3) rotation, the zeroth order eigenvalues are
\begin{align}
\bar{\lambda}^{\prime\prime}_1&=c^2_{\bar{\alpha}_{13}}\bar{\lambda}^\prime_{1}+s^2_{\bar{\alpha}_{13}}\bar{\lambda}^\prime_3+2s_{\bar{\alpha}_{13}}c_{\bar{\alpha}_{13}}c_{\bar{\alpha}_{23}}\tilde{s}_{12}\epsilon^\prime \Delta m^2_{ee}\notag\\
&\simeq \lambda_1-(\epsilon^\prime\Delta m^2_{ee})^2\frac{\tilde{s}^2_{12}}{\Delta\lambda_{31}}+\mathcal{O}(\epsilon^{\prime\,3}), \notag\\
\bar{\lambda}^{\prime\prime}_2&=\bar{\lambda}^\prime_2\simeq\lambda_2-(\epsilon^\prime\Delta m^2_{ee})^2\frac{\tilde{c}^2_{12}}{\Delta\lambda_{32}}+\mathcal{O}(\epsilon^{\prime\,3}),\notag\\
\bar{\lambda}^{\prime\prime}_3&=
s^2_{\bar{\alpha}_{13}}\bar{\lambda}^\prime_{1}+c^2_{\bar{\alpha}_{13}}\bar{\lambda}^\prime_3-2s_{\bar{\alpha}_{13}}c_{\bar{\alpha}_{13}}c_{\bar{\alpha}_{23}}\tilde{s}_{12}\epsilon^\prime \Delta m^2_{ee}\notag\\
&\simeq \lambda_3+(\epsilon^\prime\Delta m^2_{ee})^2\left(\frac{\tilde{s}^2_{12}}{\Delta\lambda_{31}}+\frac{\tilde{c}^2_{12}}{\Delta\lambda_{32}}\right)+\mathcal{O}(\epsilon^{\prime\,3}). 
\label{eq:lambdaPPbar}
\end{align}
It is easy to see that compared with Eq.~\ref{eq:lambdaPP}, $\lambda^{\prime\prime}_i\simeq\bar{\lambda}^{\prime\prime}_i$ to second order. Finally the (1-2) rotation angle is
\begin{align}
\bar{\alpha}_{12}&=\frac{1}{2}\arctan\frac{2\epsilon^\prime\Delta m_{ee}^2s_{\bar{\alpha}_{23}}c_{\bar{\alpha}_{13}}\tilde{s}_{12}}{\Delta\bar{\lambda}^{\prime\prime}_{21}}\notag\\
&\simeq\frac{(\epsilon^\prime\Delta m_{ee}^2)^2\tilde{s}_{12} \tilde{c}_{12}}{\Delta\lambda_{21}\Delta\lambda_{32}}+\mathcal{O}(\epsilon^{\prime\,4}).
\end{align}
Compared with Eq.~\ref{eq:alpha12}, now even to the leading order $\alpha_{12}\neq \bar{\alpha}_{12}$. Later we will see that this inequality is necessary for the equivalence of the eigenstates for neutrino and anti-neutrino cases. After the $U_{12}$ rotation, the corrected eigenvalues are
\begin{align}
\bar{\lambda}^{\prime\prime\prime}_1&=c^2_{\bar{\alpha}_{12}}\bar{\lambda}^{\prime\prime}_1+s^2_{\bar{\alpha}_{12}} \bar{\lambda}^{\prime\prime}_2-2s_{\bar{\alpha}_{12}}c_{\bar{\alpha}_{12}}c_{\bar{\alpha}_{13}}s_{\bar{\alpha}_{23}}\tilde{s}_{12} \epsilon^{\prime} \Delta m^2_{ee}\notag\\
&\simeq\lambda_1-(\epsilon^\prime\Delta m^2_{ee})^2\frac{\tilde{s}^2_{12}}{\Delta\lambda_{31}}+\mathcal{O}(\epsilon^{\prime\,3}), \notag\\
\bar{\lambda}^{\prime\prime\prime}_2&=s^2_{\bar{\alpha}_{12}}\bar{\lambda}^{\prime\prime}_1+c^2_{\check{\alpha}_{12}} \bar{\lambda}^{\prime\prime}_2+2s_{\bar{\alpha}_{12}}c_{\bar{\alpha}_{12}}c_{\bar{\alpha}_{13}}s_{\bar{\alpha}_{23}}\tilde{s}_{12} \epsilon^\prime \Delta m^2_{ee}\notag\\
&\simeq\lambda_2-(\epsilon^\prime\Delta m^2_{ee})^2\frac{\tilde{c}^2_{12}}{\Delta\lambda_{32}}+\mathcal{O}(\epsilon^{\prime\,3}),\notag\\
\bar{\lambda}^{\prime\prime\prime}_3&= \bar{\lambda}^{\prime\prime}_3\simeq \lambda_3+(\epsilon^\prime\Delta m^2_{ee})^2\left(\frac{\tilde{s}^2_{12}}{\Delta\lambda_{31}}+\frac{\tilde{c}^2_{12}}{\Delta\lambda_{32}}\right)+\mathcal{O}(\epsilon^{\prime\,3}). \label{eq:lambdaPPPbar}
\end{align}
By comparing the above eigenvalues after three additional rotations with the ones in the case of neutrinos, we find that $\lambda^{\prime\prime\prime}_i$ and $\bar{\lambda}^{\prime\prime\prime}_i$ are identical to second order in $\epsilon^\prime$.

\subsection{Rotated eigenstates}\label{rotatedeigenstates}
The zeroth order energy eigenstates $\ket{\check{\nu}}$ before the additional rotations are defined in Eq.~\ref{eq:zerothstates}. If
\begin{equation}
W=\begin{cases}
U_{13}(\alpha_{13})U_{23}(\alpha_{23})U_{12}(\alpha_{12}) &\text{for neutrinos} \\
U_{23}(\bar{\alpha}_{23})U_{13}(\bar{\alpha}_{13})U_{12}(\bar{\alpha}_{12}) & \text{for anti-neutrinos}
\end{cases},\label{eq:W}
\end{equation}
then the eigenstates after the rotations are
\begin{equation}
\ket{\check{\nu}}_{W}=W^{\dagger}\ket{\check{\nu}},
\end{equation}
and $U^m_{\text{PMNS}}$ from Appendix \ref{Appendixzerothorder} and Ref.~\cite{Denton:2016wmg} is corrected to be
\begin{equation}
V=U^m_\text{PMNS}W. 
\end{equation}
In the case of neutrinos, with Eqs.~\ref{eq:alpha13}, \ref{eq:alpha23}, and \ref{eq:alpha12} it is easy to verify that $U_{13}(\alpha_{13})U_{23}(\alpha_{23})U_{12}(\alpha_{12})$ can be expanded through second order to
\begin{widetext}
\begin{flalign}
U_{13}(\alpha_{13})U_{23}(\alpha_{23})U_{12}(\alpha_{12})\simeq\mathbb1+\underbrace{\epsilon^\prime\Delta m_{ee}^2\begin{pmatrix}
 & & -\frac{\tilde{s}_{12}}{\Delta\lambda_{31}} \\
 & & \frac{\tilde{c}_{12}}{\Delta\lambda_{32}} \\
\frac{\tilde{s}_{12}}{\Delta\lambda_{31}} & -\frac{\tilde{c}_{12}}{\Delta\lambda_{32}} & \\
\end{pmatrix}}_{W_1} \notag \\
\underbrace{-\frac{(\epsilon^\prime\Delta m_{ee}^2)^2}{2} \left(\begin{array}{ccc}
\left(\frac{\tilde{s}_{12}}{\Delta\lambda_{31}}\right)^2 &-\frac{2\tilde{s}_{12}\tilde{c}_{12}}{\Delta\lambda_{32}\Delta\lambda_{21}} & 0 \\
\frac{2\tilde{s}_{12}\tilde{c}_{12}}{\Delta\lambda_{31}\Delta\lambda_{21}} & \left(\frac{\tilde{c}_{12}}{\Delta\lambda_{32}}\right)^2 & 0 \\
0 & 0 & \left(\frac{\tilde{s}_{12}}{\Delta\lambda_{31}}\right)^2+\left(\frac{\tilde{c}_{12}}{\Delta\lambda_{32}}\right)^2 \\
\end{array}
\right)}_{W_2}. \label{eq:threerotations}
\end{flalign}
\end{widetext}
This expression still holds if we perform the (2-3) rotation before the (1-3) since it can be demonstrated that
\begin{multline}
U_{13}(\alpha_{13})U_{23}(\alpha_{23})U_{12}(\alpha_{12})\\
=U_{23}(\bar{\alpha}_{23})U_{13}(\bar{\alpha}_{13})U_{12}(\bar{\alpha}_{12})+\mathcal{O}(\epsilon^{\prime\,3}).
\end{multline}
Several remarkable observations in Eq.~\ref{eq:threerotations} are listed below.
\begin{itemize}
\item Both $U_{13}(\alpha_{13})$ and $U_{23}(\alpha_{23})$ contribute to the first order term $W_1$. For example, if $\alpha_{13}=0$, $(W_{1})_{13}$ and $(W_1)_{31}$ equal zero; and if $\alpha_{23}=0$, $(W_{1})_{23}$ and $(W_1)_{32}$ vanish.
\item Since $\alpha_{12}$ contributes only at second order, if we just perform the first two additional rotations, i.e.~$\alpha_{12}=0$ , the first order $W_1$ will not be affected.
\item $U_{12}(\alpha_{12})$ does contribute to the second order term $W_2$. For example, $(W_2)_{21}=0$ if $\alpha_{12}=0$. That is, although the eigenvalues after two and three additional rotations, i.e.~$\lambda^{\prime\prime}_i$ and $\lambda^{\prime\prime\prime}_i$ are identical to second order, the eigenstates are not. 
\end{itemize}
These observations are necessary to the following discussions about the relations between the additional rotations and perturbation theory. 

\subsection{Comparison with perturbation theory}\label{comparewithperturbation}
The normal approach to calculate the energy eigenvalues, eigenstates and oscillation probabilities in matter has been via a series expansion in some small parameter. For example, in \cite{Denton:2016wmg}, a three rotation approach was adopted, i.e.~performing one constant rotation $U_{23}(\theta_{23},\delta)$ followed by two rotations $U_{13}(\tilde{\theta}_{13})$ and $U_{12}(\tilde{\theta}_{12})$.
Then perturbation theory was applied wherein the eigenvalues and eigenvectors were perturbatively expanded to successive orders in $\eps'$.

\begin{figure*}[t]
\centering
\includegraphics[width=4in]{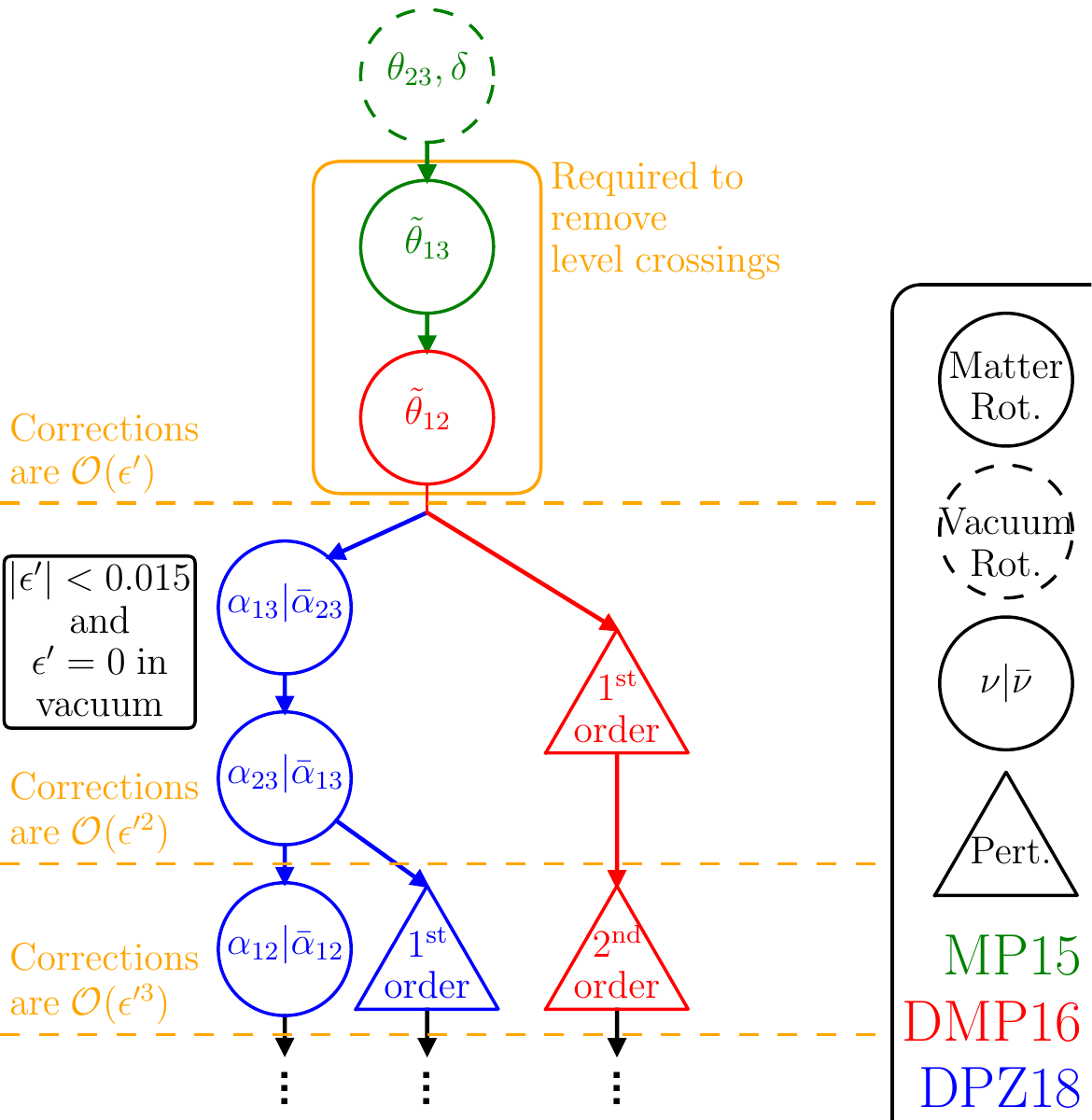}
\caption{The equivalences between the additional rotations (circles) and the perturbative expansions of the eigenvalues and eigenvectors (triangles). Performing one additional rotation is not equal to any perturbation expansion; performing two additional rotations in (1-3) and (2-3) (exchange the two for anti-neutrinos) sectors is equal to a first order perturbation expansion; performing one more additional rotation in (1-2) sector is equal to a second order perturbation expansion. The steps shown in green, red, and blue refer to Ref.~\cite{Minakata:2015gra}, Ref.~\cite{Denton:2016wmg}, and this work respectively. Another possible perturbative branch (in blue) is that if we implement a first order perturbative expansion after the $U_{13}(\alpha_{13})$ and $U_{23}(\alpha_{23})$ (or $U_{23}(\bar{\alpha}_{23})$ and $U_{13}(\bar{\alpha}_{13})$ for anti-neutrinos) rotations, the eigenvalues and eigenstates also will be corrected to ${\cal O}(\epsilon^{\prime\,2})$ accuracy, see Appendix~\ref{Appendixasideexpansion}.}
\label{fig:Schematic}
\end{figure*}

With the perturbing Hamiltonian $\check{H}_1$, we assume that by perturbation theory, the eigenstates are corrected to be
\begin{equation}
\ket{\check{\nu}}_{W^P}=W^{P\dagger}\ket{\check{\nu}}.
\end{equation}
Since $\check{H}_1$ is order $\epsilon^\prime$, we can expand $W^P$ in a series of $\epsilon^\prime$,
\begin{equation}
W^P=\mathbb{1}+W^P_1+W^P_2+\dots,
\end{equation}
and the corrected eigenvalues can also be expanded as
\begin{equation}
\lambda^P_{i}=\lambda_i+\lambda^{P(1)}_{i}+\lambda^{P(2)}_i+\dots,
\end{equation}
where $W^P_n$ and $\lambda^{P(n)}_i$ are proportional to $\epsilon^{\prime n}$, their full expressions can be found in Appendix \ref{Appendixperturbationtheory}. Comparing the results from the perturbation theory and the additional rotations, we find the following equivalences
\begin{equation}
W_1=W^P_1, \quad W_2=W^P_2, \label{eq:statesequiv}
\end{equation}
and
\begin{align}
\lambda^{\prime\prime}_i&\simeq\lambda_i+\lambda^{P(1)}_i+\mathcal{O}(\epsilon^{\prime 2})\simeq \lambda_i+\lambda^{P(1)}_i+\lambda^{P(2)}_i+\mathcal{O}(\epsilon^{\prime\,3}), \notag \\
\lambda^{\prime\prime\prime}_i&\simeq \lambda_i+\lambda^{P(1)}_i+\lambda^{P(2)}_i+\mathcal{O}(\epsilon^{\prime\,3}).\label{eq:eigenvalueequiv}
\end{align}
From Eq.~\ref{eq:statesequiv} and the observations at the end of Sec.~\ref{rotatedeigenstates}, we can make the following conclusions of the eigenstates
\begin{itemize}
\item After performing one additional rotation ($U_{13}(\alpha_{13})$ for neutrinos and $U_{23}(\bar\alpha_{23})$ for anti-neutrinos), the accuracy of the rotated eigenstates is not improved compared with the initial zeroth order $\ket{\check{\nu}}$, i.e.~errors of the eigenstates are still in $\mathcal{O}(\epsilon^\prime)$.
\item For neutrinos (anti-neutrinos), after performing two additional rotations in (1-3) and then (2-3) sectors ((2-3) and then (1-3) sectors), errors of the rotated eigenstates are diminished to $\mathcal{O}(\epsilon^{\prime\,2})$. Thus the eigenstates are equivalent to the ones of a first order perturbation theory through $\mathcal O(\eps')$ terms.
\item Errors of the eigenstates will be further diminished to $\mathcal{O}(\epsilon^{\prime\,3})$ by performing just one more rotation in (1-2) sector. Now the eigenstates have the same accuracy as the ones from a second order perturbation theory.
\end{itemize}
From Eq.~\ref{eq:eigenvalueequiv}, we can make the following conclusions of the eigenvalues
\begin{itemize}
\item Errors of the eigenvalues after the first two additional rotations are already lower than $\mathcal{O}(\epsilon^{\prime\,3})$ (that is, the eigenvalues are correct through $\mathcal{O}(\epsilon^{\prime\,2})$). To reconcile with the conclusions of the eigenstates, we say that the eigenvalues after the first two additional rotations have at least the accuracy of the first order perturbation theory.
\item Errors of the eigenvalues after the three additional rotations are even smaller, so of course lower than $\mathcal{O}(\epsilon^{\prime\,3})$. Again to reconcile with the conclusions of the eigenstates, we say that their accuracy is at least equivalent to the ones corrected by a second order perturbation theory.
\end{itemize}
Now we combine the conclusions of the eigenvalues and the eigenstates. We find two equivalences between the additional rotations and the perturbation theory. 
\begin{itemize}
\item By performing two additional rotations in (1-3) and (2-3) sector (the order is exchanged for anti-neutrinos), we can improve the eigenstates and eigenvalues to be as precise as the ones from first order perturbation theory. 
\item By performing three additional rotations, we can improve the eigenstates and eigenvalues to be as precise as the ones from a second order perturbation theory.
\end{itemize}
All the conclusions are also summarized in Fig.~\ref{fig:Schematic}.

\section{Corrections to the mixing angles and the CP phase}
\label{sec:corrections to angles}
After the three additional rotations, the corrected PMNS matrix in matter is $V=U^m_\text{PMNS}W$. Since $W$ is a real special orthogonal matrix, $V$ can be written as 
\begin{align}
V&=e^{iA}U_{23}(\tilde{\theta}^\prime_{23},\tilde{\delta}^\prime)U_{13}(\tilde{\theta}^\prime_{13})U_{12}(\tilde{\theta}^\prime_{12})e^{iB}\notag\\
&=U_{23}(\tilde{\theta}_{23},\tilde{\delta})U_{13}(\tilde{\theta}_{13})U_{12}({\tilde{\theta}_{12}})W, \label{eq:V}
\end{align}
Here $A$ and $B$ are some real diagonal matrices. In general, $A$ and $B$ are necessary to get real solutions of $\tilde{\theta}^\prime_{ij}$ and $\tilde{\delta}^\prime$. Since both $A$ and $B$ are real and diagonal, they only add some additional complex phases to the eigenstates, which will not change any physics. 

We can expand $\tilde{\theta}^\prime_{ij}$ as
\begin{equation}
\tilde{\theta}^\prime_{ij}\simeq \tilde{\theta}_{ij}+\tilde{\theta}^{(1)}_{ij}+\tilde{\theta}^{(2)}_{ij}+..., \label{eq:thetaexpansion}
\end{equation}
where $\tilde{\theta}^{(n)}_{ij}$ is proportional to $\epsilon^{\prime\,n}$.   

To first order, $W=U_{13}(\alpha_{13})U_{23}(\alpha_{23})=\mathbb{1}+W_1+\mathcal{O}(\epsilon^{\prime\,2})$. Details of $W_1$ can be found in Eq.~\ref{eq:threerotations}. We give the final results here. The first order corrections to the mixing angles and CP phase are
\begin{align}
\tilde{\theta}^{\,(1)}_{13}={}&\epsilon^\prime\Delta m_{ee}^2 \tilde{s}_{12}\tilde{c}_{12} \left(\frac{1}{\Delta\lambda_{32}}-\frac{1}{\Delta\lambda_{31}}\right), \notag \\
\tilde{\theta}^{\,(1)}_{12}={}& -\epsilon^\prime\Delta m_{ee}^2\frac{\tilde{s}_{13}}{\tilde{c}_{13}} \left(\frac{\tilde{s}^2_{12}}{\Delta\lambda_{31}}+\frac{\tilde{c}^2_{12}}{\Delta\lambda_{32}}\right), \notag \\
\tilde{\theta}^{\,(1)}_{23}={}& \epsilon^\prime\Delta m_{ee}^2 \frac{\tilde{c}_\delta}{\tilde{c}_{13}} \left(\frac{\tilde{s}^2_{12}}{\Delta\lambda_{31}}+\frac{\tilde{c}^2_{12}}{\Delta\lambda_{32}}\right), \notag \\
\tilde{\delta}^{\,(1)}={}& -\epsilon^\prime\Delta m_{ee}^2\frac{2c_{2\tilde{\theta}_{23}}\tilde{s}_\delta}{s_{2\tilde{\theta}_{23}}\tilde{c}_{13}} \left(\frac{\tilde{s}^2_{12}}{\Delta\lambda_{31}}+\frac{\tilde{c}^2_{12}}{\Delta\lambda_{32}}\right).
\end{align}

Please note that since $e^{ip}W_1 e^{-ip}=W_1$ for any real number $p$, it's free to set one of the diagonal elements of $A$ or $B$ to be zero. 
All the corrections to the mixing angles and the CP phase are invariants under a transformation of exchanging $\lambda_1$, $\lambda_2$ and $\tilde{\theta}_{12}\Rightarrow\tilde{\theta}_{12}\pm\frac{\pi}{2}$. This is easy to verify in the above equations. More details can be found in Appendix \ref{Appendixsymmetry}.
 
To second order, $W=U_{13}(\alpha_{13})U_{23}(\alpha_{23})U_{12}(\alpha_{12})=\mathbb{1}+W_1+W_2+\mathcal{O}(\epsilon^{\prime\,3})$, combining with the first order results the second order perturbations can be solved. Details of the second order results are listed in Appendix \ref{Appendixsecondordercorrections}. 
The corrected mixing angles and CP phase through second order are
\begin{align}
\tilde{s}^{\,\prime}_{13}\simeq{}&\tilde{s}_{13}+\epsilon^\prime\Delta m^2_{ee} \tilde{s}_{12} \tilde{c}_{12}\tilde{c}_{13}\left(\frac{1}{\Delta\lambda_{32}}-\frac{1}{\Delta\lambda_{31}}\right)+f^{(2)}_{13}, \notag \\
\tilde{s}^{\,\prime}_{12}\simeq{}&\tilde{s}_{12}-\epsilon^\prime\Delta m^2_{ee}\frac{\tilde{s}_{13} \tilde{c}_{12}}{\tilde{c}_{13}}\left(\frac{\tilde{s}^2_{12}}{\Delta\lambda_{31}}+\frac{\tilde{c}^2_{12}}{\Delta\lambda_{32}}\right)+f^{(2)}_{12} ,\notag \\
\tilde{s}^{\,\prime}_{23}\simeq{}&\tilde{s}_{23}+\epsilon^\prime\Delta m^2_{ee}\frac{\tilde{c}_\delta \tilde{c}_{23}}{\tilde{c}_{13}}\left(\frac{\tilde{s}^2_{12}}{\Delta\lambda_{31}}+\frac{\tilde{c}^2_{12}}{\Delta\lambda_{32}}\right)+f^{(2)}_{23}
, \notag \\
\tilde{s}^{\,\prime}_{\delta}\simeq{}&\tilde{s}_\delta-\epsilon^\prime\Delta m^2_{ee}\frac{c_{2\tilde{\theta}_{23}}s_{2\tilde{\delta}}}{s_{2\tilde{\theta}_{23}}\tilde{c}_{13}}\left(\frac{\tilde{s}^2_{12}}{\Delta\lambda_{31}}+\frac{\tilde{c}^2_{12}}{\Delta\lambda_{32}}\right)+f^{(2)}_{\delta} \label{eq:sP}.
\end{align}
Functions of the second order terms $f^{(2)}$, which are proportional to $\epsilon^{\prime\,2}$, can be found in the Appendix \ref{Appendixsecondordercorrections}. 

\subsection{Numerical tests}
\begin{figure*}
\centering
\includegraphics[scale=0.55]{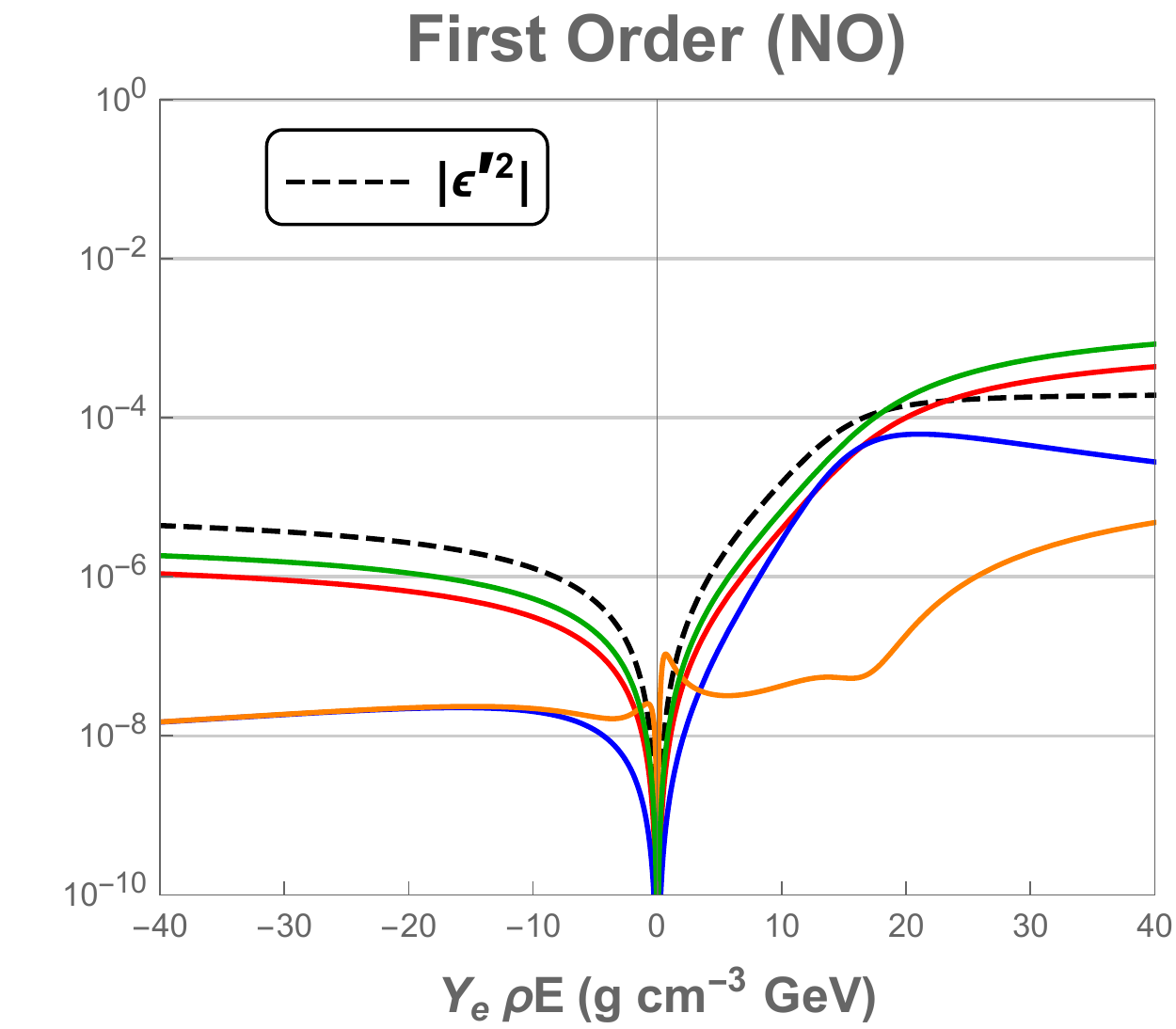}
\includegraphics[scale=0.55]{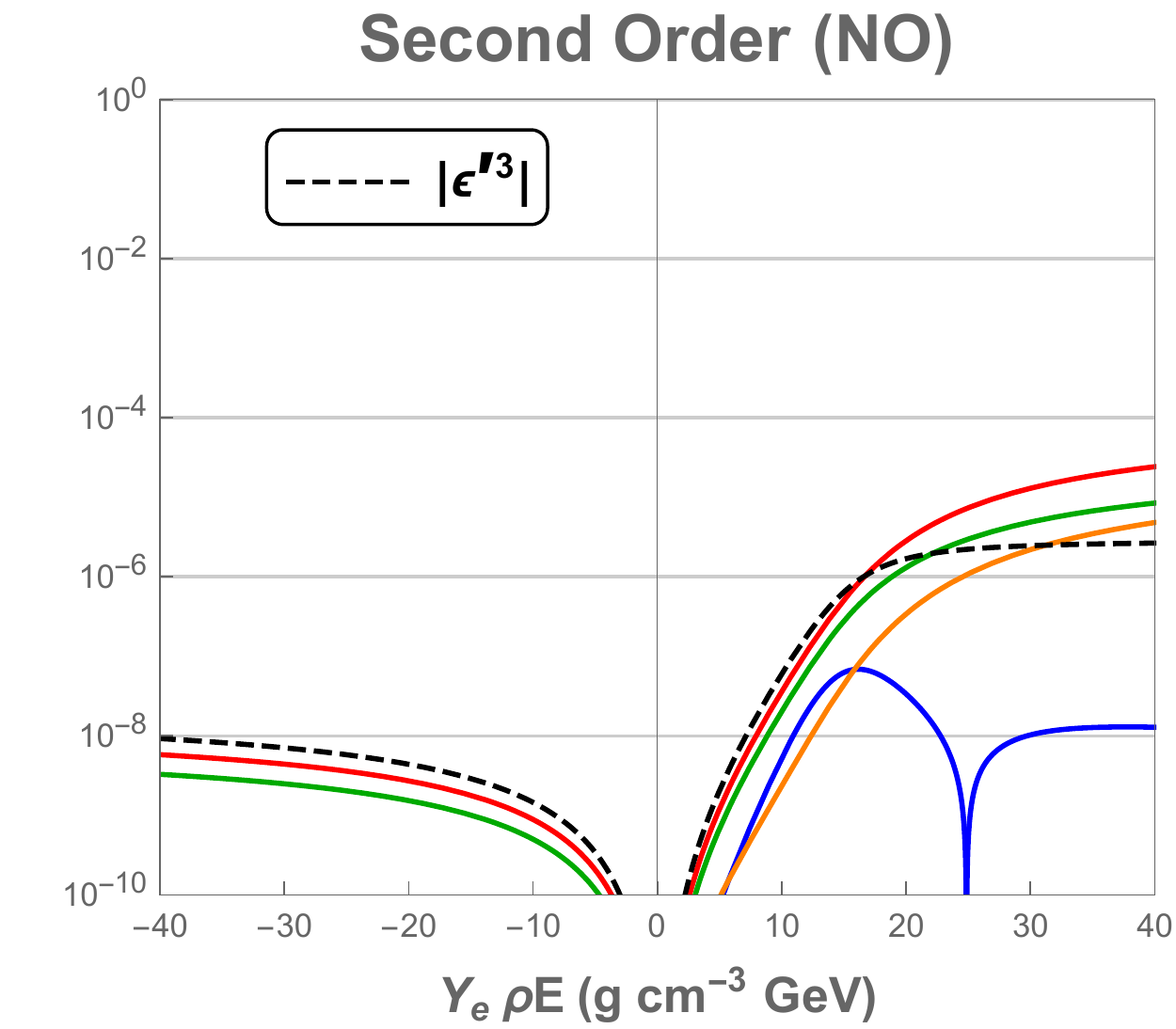}

\vspace{0.5cm}

\qquad\includegraphics[scale=0.55]{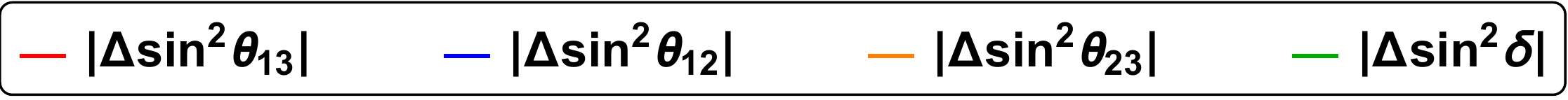}
\caption{The absolute accuracy of the approximations of the mixing angles and CP phase in matter in this paper to first order (left) and second order (right) for the normal mass ordering. The black dashed curves in the left and right plots are $|\epsilon^{\prime2}|$ and $|\epsilon^{\prime 3}|$, respectively. It is evident that the error of $\sin^2$ of each mixing angle and phase at first (second) order is about $\eps^{\prime2}$ ($\epsilon^{\prime3}$).}
\label{fig:d1_angles}
\end{figure*}

Neutrino propagation in constant density matter has been analytically studied, the accurate mixing angles and CP phase can be found in \cite{Barger:1980tf,Zaglauer:1988gz}. Our formulas have second order accuracy so it is expected that the differences between the analytical solutions and our approximations are significantly below $\epsilon^{\prime2}$ and even to first order there are precise to $>10^{-3}$. We show the precision of the angles to first and second order in Fig.~\ref{fig:d1_angles} for the normal mass ordering. It is evident that the approximated values achieve the expected accuracy.

\section{Conclusions}
\label{sec:conclusions}
We have significantly improved the accuracy and understanding of the recent perturbative framework for neutrino propagations in uniform matter in \cite{Denton:2016wmg}. This has been achieved by performing additional rotations which diagonalize the sectors with leading order off-diagonal elements of the Hamiltonian. The primary advantage of this approach is that the zeroth order Hamiltonian is applicable to the whole range of matter potential $a$, whereas perturbation expansions are most reliable for weak matter effect. By studying orders of the off-diagonal elements of the perturbing Hamiltonian, we determine the sequence of the additional rotations. For neutrinos the sequence is $U_{13}\Rightarrow U_{23}\Rightarrow U_{12}$, and for anti-neutrinos $U_{13}$, $U_{23}$ are exchanged. The additional rotation angles are solved to diagonalize the corresponding sectors. The first two rotation angles in (1-3) and (2-3) sectors have first order (in $\epsilon^\prime$) whereas the third angle in (1-2) sector is second order. The diagonal elements of the rotated Hamiltonian, which are the approximations to the eigenvalues, are calculated to second order.

We compare the eigenvalues and eigenstates derived by the additional rotations and perturbation theories and reveal the equivalences. Performing two successive additional rotations in (1-3) and (2-3) sectors is equal to a first order perturbation theory. Performing three successive additional rotations in (1-3), (2-3) and (1-2) sectors is equal to a second order perturbative expansion.

Finally, we derive first order approximation formulas of the mixing angles and CP phase in matter and compare them with the exact solutions. Numerical tests show that regardless the scale of matter potential, errors of the first order approximation formulas are controlled to be no more than $10^{-5}$, achieving the expected accuracy with a significant computational speed improvement as well \cite{Kelly:2018kmb}. More precise approximations to second order are given in Appendix~\ref{Appendixsecondordercorrections}.

\section{Acknowledgement}
This manuscript has been authored by Fermi Research Alliance, LLC under Contract No.
DE-AC02-07CH11359 with the U.S. Department of Energy, Office of Science, Office of High
Energy Physics.

SP thanks IFT of Madrid for wonderful hospitality during part of this work.
This project has received funding/support from the European Union’s Horizon 2020
research and innovation programme under the Marie Sklodowska-Curie grant agreement
No 690575.  This project has received funding/support from the European Union’s Horizon 
2020 research and innovation programme under the  Marie Sklodowska-Curie grant
agreement No 674896.

PBD acknowledges support from the Villum Foundation (Project No. 13164) and the
Danish National Research Foundation (DNRF91 and Grant No. 1041811001).

\appendix
\section{Zeroth order eigenvalues and mixing angles} \label{Appendixzerothorder}
The derivation process of the mixing angles and Eq.~\ref{eq:Hbar} is presented in this Appendix.
\subsection{\texorpdfstring{$U_{23}(\tilde{\theta}_{23},\tilde{\delta})$}{U23(tilde theta23, tilde delta)} rotation}
Define
\begin{equation}
\tilde{H}\equiv U^\dagger_{23}(\tilde{\theta}_{23},\tilde{\delta})HU_{23}(\tilde{\theta}_{23},\tilde{\delta}).
\end{equation}
Now $\tilde{H}$ is real and does not depend on $\theta_{23}$ and $\delta$. 
\begin{align}
\tilde{H}={}&\frac{1}{2E}\left(\begin{array}{ccc}
\lambda_a & & s_{13}c_{13}\Delta m^2_{ee}\\
& \lambda_b & \\
s_{13}c_{13}\Delta m^2_{ee}& & \lambda_c \\
\end{array}
\right)\notag\\
&+\epsilon s_{12}c_{12}\frac{\Delta m^2_{ee}}{2E}\left(\begin{array}{ccc}
&c_{13} & \\
c_{13}& &-s_{13} \\
&-s_{13} & \\
\end{array}
\right),
\end{align}
where
\begin{align}
\lambda_a&=a+(s^2_{13}+\epsilon s^2_{12})\Delta m^2_{ee}, \notag \\
\lambda_b&=\epsilon c^2_{12}\Delta m^2_{ee}, \notag \\
\lambda_c&=(c^2_{13}+\epsilon s^2_{12})\Delta m^2_{ee}.
\end{align}

\subsection{\texorpdfstring{$U_{13}(\tilde{\theta}_{13})$}{U13(tilde theta13)} rotation}
Observe the entries of $\tilde{H}$, it's easy to see that the (1-3) sector contributes the leading order off-diagonal entries. Therefore it's reasonable to make $U_{13}(\tilde{\theta}_{13})$ diagonalize this sector. After this rotation
\begin{align}
\hat{H}\equiv{}&U^\dagger_{13}(\tilde{\theta}_{13})\tilde{H}U_{13}(\tilde{\theta}_{13}) \notag \\
={}&\frac{1}{2E}\left(\begin{array}{ccc}
\lambda_-& & \\
&\lambda_0 & \\
& &\lambda_+ \\
\end{array}
\right)+\epsilon s_{12}c_{12}\frac{\Delta m^2_{ee}}{2E}\notag\\
&\times\left(\begin{array}{ccc}
&c_{(\tilde{\theta}_{13}-\theta_{13})} & \\
c_{(\tilde{\theta}_{13}-\theta_{13})}& & s_{(\tilde{\theta}_{13}-\theta_{13})}\\
&s_{(\tilde{\theta}_{13}-\theta_{13})} & \\
\end{array}
\right), \label{eq:Hhat}
\end{align} 
where
\begin{align}
\lambda_\pm={}&\frac{1}{2}\left[(\lambda_a+\lambda_c)\right.\notag\\
&\left.\pm\text{sign}(\Delta m^2_{ee})\sqrt{(\lambda_a-\lambda_c)^2+4(s_{13}c_{13}\Delta m^2_{ee})^2}\right], \notag \\
\lambda_0={}&\epsilon c^2_{12}\Delta m^2_{ee}.
\end{align}
With the diagonal elements above, $\tilde{\theta}_{13}$ can be determined by
\begin{equation}
\sin^2\tilde{\theta}_{13}=\frac{\lambda_+-\lambda_c}{\lambda_+-\lambda_-}, \quad \tilde{\theta}_{13}\in [0,\pi/2].
\end{equation}

\subsection{\texorpdfstring{$U_{12}(\tilde{\theta}_{12})$}{U12(tilde theta12)} rotation}
For any long baseline experiment the largest off diagonal terms are in the (1-2) sector (see subsection \ref{ssec:t12vt23} below for a caveat).
Now $U_{12}(\tilde{\theta}_{12})$ is required to diagonalize the (1-2) sector of $\hat{H}$, and $\check{H}$ is obtained after is rotation.
\begin{align}
\check{H}={}&U^\dagger_{12}(\tilde{\theta}_{12})\hat{H}U_{12}(\tilde{\theta}_{12}) \notag \\
={}&\frac{1}{2E}\left(\begin{array}{ccc}
\lambda_1& & \\
&\lambda_2 & \\
& &\lambda_3 \\
\end{array}
\right)\notag\\
&+\epsilon s_{(\tilde{\theta}_{13}-\theta_{13})}s_{12}c_{12}\frac{\Delta m^2_{ee}}{2E}\left(\begin{array}{ccc}
& &-\tilde{s}_{12} \\
& & \tilde{c}_{12}\\
-\tilde{s}_{12} & \tilde{c}_{12} & \\
\end{array}
\right) ,
\end{align}
where
\begin{align}
\lambda_{1,2}={}&\frac{1}{2}\left[(\lambda_0+\lambda_-)\right.\notag\\
&\left.\mp \sqrt{(\lambda_0-\lambda_-)^2+4(\epsilon c_{(\tilde{\theta}_{13}-\theta_{13})}s_{12}c_{12}\Delta m^2_{ee})^2}\right], \notag \\
\lambda_3={}&\lambda_+, \label{eq:lambdapm0}
\end{align}
and
\begin{equation}
\sin^2\tilde{\theta}_{12}=\frac{\lambda_2-\lambda_0}{\lambda_2-\lambda_1}, \quad \tilde{\theta}_{12}\in [0,\pi/2].
\end{equation}

Alternative ways to write these expressions can be found in~\cite{Denton:2018hal}.

\subsection{$\tilde{\theta}_{12}$ vs.~$\acute\theta_{23}$}
\label{ssec:t12vt23}
After the $(\tilde{\theta}_{23},\tilde{\delta})$ and $\tilde{\theta}_{13}$ rotations, the Hamiltonian is given by Eq.~\ref{eq:Hhat}.

If we follow the simplest prescription of rotating away the largest off diagonal elements as we have for the previous steps, we perform the rotation in the (1-2) sector, which also removes the solar level crossing and returns the PMNS order.
This is the largest off-diagonal terms when $\c>\s$ which is valid for neutrinos with $E<11.5$ GeV ($a<\dmsqee/c_{2\theta_{13}}$)\footnote{Note that the threshold is slightly higher than the atmospheric resonance at $a=\dmsqee c_{2\theta_{13}}$.} and for all anti-neutrinos in the NO.
Thus in the NO for the neutrinos above the atmospheric resonance (or anti-neutrinos above the atmospheric resonance in the IO) it is better to diagonalize the (2-3) sector next.
While this does not address the level-crossing at the solar resonance, it is immaterial since we are focusing on neutrinos with $E>11.5$ GeV.

For the case $E>11.5$ GeV, the new mixing angle denoted $\acute\theta_{23}$ is given by
\begin{equation}
\tan2\acute\theta_{23}=\frac{2\eps s_{12}c_{12}\s\dmsqee}{\Dl+0}\,,
\end{equation}
and the eigenvalues $\lambda_{x,y,z}$ are
\begin{align}
\lambda_x&=\lambda_-\,,\nonumber\\
\lambda_y&=c_{\acute\theta_{23}}^2\lambda_0+s_{\acute\theta_{23}}^2\lambda_+-2\eps s_{\acute\theta_{23}}c_{\acute\theta_{23}}s_{12}c_{12}\s\dmsqee\,, \nonumber\\
\lambda_z&=s_{\acute\theta_{23}}^2\lambda_0+c_{\acute\theta_{23}}^2\lambda_++2\eps s_{\acute\theta_{23}}c_{\acute\theta_{23}}s_{12}c_{12}\s\dmsqee\,. \nonumber \\
\end{align}
The new perturbing Hamiltonian is
\begin{equation}\eps\c s_{12}c_{12}\dmsqee
\begin{pmatrix}
&c_{\acute\theta_{23}}&-s_{\acute\theta_{23}}\\
c_{\acute\theta_{23}}\\
-s_{\acute\theta_{23}}
\end{pmatrix}\,.
\end{equation}
In general we will assume that $E<11.5$ GeV and use the $\tilde\theta_{12}$ rotation since it also addresses the level crossing and matches the PMNS order.
In addition, the difference between $\c$ and $\s$ is small until well past the atmospheric resonance.

\section{Hamiltonians after the additional rotations}\label{Appendixadditionalrotations}

\begin{table*}[t]
\begin{tabular}{c}
\centering
\large\textbf{Neutrinos}
\end{tabular}

\vspace{0.3cm}

\begin{tabular}{c|c|ccc|c}
Rotation angles\,&$\,2E\,H_0$  & $2E\,(H_1)_{12}/\mathcal{N}\quad$ &  $2E\,(H_1)_{13}/\mathcal{N}\quad$   &  $2E\,(H_1)_{23}/\mathcal{N}\,$& $\mathcal{N}$  \\ \hline
 & $(\lambda_a,\,\lambda_b,\,\lambda_c)$   & $c_{13} \,s_{12} c_{12} \epsilon$  & $s_{13} c_{13}$& $s_{13} \,s_{12} c_{12} \epsilon$ & $\Delta m^2_{ee}$  \\
 $\tilde{\theta}_{13}$ &  $(\lambda_-,\,\lambda_0,\,\lambda_+)$  &  $c_{(\tilde{\theta}_{13}-\theta_{13})}$ &  0
&$  s_{(\tilde{\theta}_{13}-\theta_{13})}  $ & $ \times\,s_{12} c_{12} \, \epsilon$  \\
 $\tilde{\theta}_{12}$ &  $(\lambda_1,\,\lambda_2,\,\lambda_3)$  &  0 &  $-\tilde{s}_{12}$
&$\tilde{c}_{12}$   & $  \times\,s_{(\tilde{\theta}_{13}-\theta_{13})}$  \\[2mm]
\hline
$\alpha_{13}$ & $(\lambda^\prime_1,\,\lambda^\prime_2,\,\lambda^\prime_3)$  & $ -s_{\alpha_{13}}$  & $0$ & $ c_{\alpha_{13}}$  & 
$ \times\,\tilde{c}_{12}$ \\
$\alpha_{23}$ & $\,(\lambda^{\prime\prime}_1,\,\lambda^{\prime\prime}_2,\,\lambda^{\prime\prime}_3)$   & 
$ c_{\alpha_{23}}$ & $s_{\alpha_{23}}$ & $0$ & $  \times\,(-s_{\alpha_{13}}) $ \\
$\alpha_{12}$ & $(\lambda^{\prime\prime\prime}_1,\,\lambda^{\prime\prime\prime}_2,\,\lambda^{\prime\prime\prime}_3)$ &  $0$ & $c_{\alpha_{12}}$  & 
$ s_{\alpha_{12}}$  & 
$ \times\,s_{\alpha_{23}} $ 
\end{tabular}

\vspace{0.5cm}

\begin{tabular}{c}
\centering
\large
\textbf{Anti-Neutrinos}
\end{tabular}

\vspace{0.3cm}

\begin{tabular}{c|c|ccc|c}
Rotation angles\,&$\,2E\,H_0$  & $2E\,(H_1)_{12}/\mathcal{N}\quad$ &  $2E\,(H_1)_{13}/\mathcal{N}\quad$  &  $2E\,(H_1)_{23}/\mathcal{N}\,$& $\mathcal{N}$  \\ \hline
 & $(\lambda_a,\,\lambda_b,\,\lambda_c)$   & $c_{13} \,s_{12} c_{12} \epsilon$ & $s_{13} c_{13}$ & $s_{13} \,s_{12} c_{12} \epsilon$ & $\Delta m^2_{ee}$  \\
 $\tilde{\theta}_{13}$ &  $(\lambda_-,\,\lambda_0,\,\lambda_+)$ &  $c_{(\tilde{\theta}_{13}-\theta_{13})}$&  0 
&$  s_{(\tilde{\theta}_{13}-\theta_{13})}  $ & $ \times\,s_{12} c_{12} \, \epsilon$  \\
 $\tilde{\theta}_{12}$ &  $(\lambda_1,\,\lambda_2,\,\lambda_3)$  &  0  &  $-\tilde{s}_{12}$ &
$\tilde{c}_{12}$   & $  \times\,s_{(\tilde{\theta}_{13}-\theta_{13})}$  \\[2mm] 
\hline
$\bar{\alpha}_{23}$ & $(\bar{\lambda}^\prime_1,\,\bar{\lambda}^\prime_2,\,\bar{\lambda}^\prime_3)$   & $-s_{\bar{\alpha}_{23}}$  & $c_{\bar{\alpha}_{23}}$ & $ 0$  & 
$ \times\,(-\tilde{s}_{12})$ \\
$\bar{\alpha}_{13}$ & $(\bar{\lambda}^{\prime\prime}_1,\,\bar{\lambda}^{\prime\prime}_2,\,\bar{\lambda}^{\prime\prime}_3)$  & $ c_{\bar{\alpha}_{13}}$   & $0$&
$s_{\bar{\alpha}_{13}}$ & 
$  \times\,(-s_{\bar{\alpha}_{23}}) $ \\
$\bar{\alpha}_{12}$ & $(\bar{\lambda}^{\prime\prime\prime}_1,\,\bar{\lambda}^{\prime\prime\prime}_2,\,\bar{\lambda}^{\prime\prime\prime}_3)$   & $0$ & $-s_{\bar{\alpha}_{12}}$ & 
$ c_{\bar{\alpha}_{12}}$  & 
$ \times\,s_{\bar{\alpha}_{13}} $ 
\end{tabular}

\vspace{0.3cm}

\caption{Entries of the Hamiltonian after each rotation for neutrinos and anti-neutrinos are presented. $\mathcal{N}$ in the last column is a normalization factor. For each row, $\mathcal{N}$ is equal to the product of all elements on and above this line. The first three rows are identical for neutrinos and anti-neutrinos.}
\label{tab:H_rotations}
\end{table*}

In the case of neutrinos, after the $U_{13}({\alpha_{13}})$ rotation, the Hamiltonian becomes $\check{H}^\prime=\check{H}^\prime_0+\check{H}^\prime_1$ where
\begin{gather}
\begin{aligned}
2E(\check{H}^\prime_0)_{11}={}&c^2_{\alpha_{13}}\lambda_1+s^2_{\alpha_{13}} \lambda_3+2s_{\alpha_{13}}c_{\alpha_{13}}\,\tilde{s}_{12}\epsilon^\prime \Delta m^2_{ee},\\
2E(\check{H}^\prime_0)_{12}={}&0,\\
2E(\check{H}^\prime_0)_{13}={}&-s_{\alpha_{13}}c_{\alpha_{13}}\Delta\lambda_{31}+(s^2_{\alpha_{13}}-c^2_{\alpha_{13}})\,\tilde{s}_{12}\epsilon^\prime\Delta m^2_{ee},\\
2E(\check{H}^\prime_0)_{22}={}&\lambda_2,\\
2E(\check{H}^\prime_0)_{23}={}&0,\\
2E(\check{H}^\prime_0)_{33}={}&s^2_{\alpha_{13}}\lambda_1+c^2_{\alpha_{13}}\lambda_3-2s_{\alpha_{13}}c_{\alpha_{13}}\,\tilde{s}_{12} \epsilon^\prime \Delta m^2_{ee}, 
\end{aligned}
\label{eq:checkH0p}
\end{gather}
and $(\check{H}^\prime_0)_{ij}=(\check{H}^\prime_0)_{ji}$, and
\begin{equation}
\check{H}^\prime_1=\frac{\epsilon^\prime\Delta m^2_{ee}\tilde{c}_{12}}{2E}
\begin{pmatrix}
& -s_{\alpha_{13}} & \\
 -s_{\alpha_{13}} & & c_{\alpha_{13}} \\
 & c_{\alpha_{13}} & \\
\end{pmatrix}
\label{eq:checkH1p}
\end{equation}
We require the (1-3) sector to be diagonalized, i.e $\alpha_{13}$ must satisfy an equation:
\begin{equation}
-s_{\alpha_{13}}c_{\alpha_{13}}\Delta\lambda_{31}+(s^2_{\alpha_{13}}-c^2_{\alpha_{13}})\,\tilde{s}_{12}\epsilon^\prime\Delta m^2_{ee}=0.
\end{equation}
The solution is Eq.~\ref{eq:alpha13}.

After the $U_{23}(\alpha_{23})$ rotation, the Hamiltonian is $\check{H}^{\prime\prime}=\check{H}^{\prime\prime}_0+\check{H}^{\prime\prime}_1$, where
\begin{gather}
\begin{aligned}
2E(\check{H}^{\prime\prime}_0)_{11}={}&\lambda^{\prime}_1,\\
2E(\check{H}^{\prime\prime}_0)_{12}={}&0,\\
2E(\check{H}^{\prime\prime}_0)_{13}={}&0,\\
2E(\check{H}^{\prime\prime}_0)_{22}={}&c^2_{\alpha_{23}}\lambda^\prime_2+s^2_{\alpha_{23}} \lambda^\prime_3\\
&-2s_{\alpha_{23}}c_{\alpha_{23}}\,c_{\alpha_{13}}\tilde{c}_{12} \epsilon^\prime \Delta m^2_{ee},\\
2E(\check{H}^{\prime\prime}_0)_{23}={}&-s_{\alpha_{23}}c_{\alpha_{23}}\Delta\lambda^\prime_{32}\\
&-(s^2_{\alpha_{23}}-c^2_{\alpha_{23}})\,c_{\alpha_{13}}\tilde{c}_{12}\epsilon^\prime\Delta m^2_{ee},\\
2E(\check{H}^{\prime\prime}_0)_{33}={}&s^2_{\alpha_{23}}\lambda^\prime_2+c^2_{\alpha_{23}} \lambda^\prime_3\\
&+2s_{\alpha_{23}}c_{\alpha_{23}}\,c_{\alpha_{13}}\tilde{c}_{12} \epsilon^\prime \Delta m^2_{ee}
\end{aligned} \label{eq:checkH0pp}
\end{gather}
and $(\check{H}^{\prime\prime}_0)_{ij}=(\check{H}^{\prime\prime}_0)_{ji}$, and
\begin{equation}
\check{H}^{\prime\prime}_1=-\frac{\epsilon^\prime\Delta m^2_{ee}\tilde{c}_{12} s_{\alpha_{13}}}{2E}
\begin{pmatrix}
& c_{\alpha_{23}} &s_{\alpha_{23}} \\
 c_{\alpha_{23}} & & \\
s_{\alpha_{23}}& & \\ 
\end{pmatrix}.
\label{eq:checkH1pp}
\end{equation}
Now the (2-3) sector must be diagonalized, i.e. $\alpha_{23}$ must satisfy 
\begin{equation}
-s_{\alpha_{23}}c_{\alpha_{23}}\Delta\lambda^\prime_{32}-(s^2_{\alpha_{23}}-c^2_{\alpha_{23}}) c_{\alpha_{13}}\tilde{c}_{12}\epsilon^\prime\Delta m^2_{ee}=0.
\end{equation}
The solution is Eq.~\ref{eq:alpha23}. Since $\alpha_{13}$ is a first order (in $\epsilon^\prime$) term, it is evident that $\check{H}_1^{\prime\prime}$ is in second order.

After the $U_{12}(\alpha_{12})$ rotation, the Hamiltonian is $\check{H}^{\prime\prime\prime}=\check{H}^{\prime\prime\prime}_0+\check{H}^{\prime\prime\prime}_1$, where
\begin{gather}
\begin{aligned}
2E(\check{H}^{\prime\prime\prime}_0)_{11}={}&c^2_{\alpha_{12}}\lambda^{\prime\prime}_1+s^2_{\alpha_{12}} \lambda^{\prime\prime}_2\\
&+2s_{\alpha_{12}}c_{\alpha_{12}}c_{\alpha_{23}}s_{\alpha_{13}}\tilde{c}_{12} \epsilon^\prime \Delta m^2_{ee},\\
2E(\check{H}^{\prime\prime\prime}_0)_{12}={}&-s_{\alpha_{12}}c_{\alpha_{12}}\Delta\lambda^{\prime\prime}_{21}\\
&+(s^2_{\alpha_{12}}-c^2_{\alpha_{12}})c_{\alpha_{23}}s_{\alpha_{13}}\tilde{c}_{12} \epsilon^\prime \Delta m^2_{ee},\\
2E(\check{H}^{\prime\prime\prime}_0)_{13}={}&0,\\
2E(\check{H}^{\prime\prime\prime}_0)_{22}={}&s^2_{\alpha_{12}}\lambda^{\prime\prime}_1+c^2_{\alpha_{12}} \lambda^{\prime\prime}_2\\
&-2s_{\alpha_{12}}c_{\alpha_{12}}c_{\alpha_{23}}s_{\alpha_{13}}\tilde{c}_{12} \epsilon^\prime \Delta m^2_{ee},\\
2E(\check{H}^{\prime\prime\prime}_0)_{23}={}&0,\\
2E(\check{H}^{\prime\prime\prime}_0)_{33}={}&\lambda^{\prime\prime}_3,
\end{aligned}
\label{eq:checkH0ppp}
\end{gather}
and $(\check{H}^{\prime\prime\prime}_0)_{ij}=(\check{H}^{\prime\prime\prime}_0)_{ji}$, and
\begin{equation}
\check{H}^{\prime\prime\prime}_1=-\frac{\epsilon^\prime\Delta m^2_{ee}\tilde{c}_{12}s_{\alpha_{13}}s_{\alpha_{23}}}{2E}\begin{pmatrix}
& & c_{\alpha_{12}} \\
 & &s_{\alpha_{12}} \\
c_{\alpha_{12}}&s_{\alpha_{12}} & \\
\end{pmatrix}.
\label{eq:checkH1ppp}
\end{equation}
It is easy to verify that $\check{H}^{\prime\prime\prime}_1$ is already a third order term in $\epsilon^\prime$ and $\alpha_{12}$ must diagonalize the (1-2) sector, i.e.
\begin{equation}
-s_{\alpha_{12}}c_{\alpha_{12}}\Delta\lambda^{\prime\prime}_{21}+(s^2_{\alpha_{12}}-c^2_{\alpha_{12}})c_{\alpha_{23}}s_{\alpha_{13}}\tilde{c}_{12} \epsilon^\prime \Delta m^2_{ee}=0.
\end{equation}
The solution is Eq.~\ref{eq:alpha12}.

The approach for anti-neutrinos is quite similar so we will not provide the detailed procedure. Alternatively we simply describe it by citing Eq.~\ref{eq:generaldiagonalize}. The first additional rotation diagonalizes the (2-3) submatrix with $\theta=\bar{\alpha}_{23}$, and $\lambda_x=\tilde{c}_{12}\epsilon^\prime\Delta m^2_{ee}$; the second additional rotation diagonalizes the (1-3) submatrix with $\theta=\bar{\alpha}_{13}$, and $\lambda_x=-c_{\bar{\alpha}_{23}}\tilde{s}_{12}\epsilon^\prime\Delta m^2_{ee}$; the third additional rotation diagonalizes the (1-2) submatrix with $\theta=\bar{\alpha}_{12}$, and $\lambda_x=c_{\bar{\alpha}_{13}}s_{\bar{\alpha}_{23}}\tilde{s}_{12}\epsilon^\prime\Delta m^2_{ee}$.   

For both cases of neutrino and anti-neutrino, the Hamiltonian after each rotation is summarized in Table~\ref{tab:H_rotations}.

\section{Perturbation expansions}
\subsection{The perturbative expansion of DMP}
\label{Appendixperturbationtheory}
Here we describe the perturbative expansions calculated from the initial zeroth order expressions from DMP \cite{Denton:2016wmg}.
By the first order perturbation theory, since all diagonal elements of $\check{H}_1$ vanish the diagonal elements of $W^P_1$ also vanish. The non-diagonal elements are
\begin{equation}
(W^P_1)_{ij}=-\frac{2E(\check{H}_1)_{ij}}{\Delta\lambda_{ij}},
\end{equation}
and from Eq.~\ref{eq:Hbar} it is easy to get
\begin{equation}
W^P_1=\epsilon^\prime\Delta m_{ee}^2\left(\begin{array}{ccc}
 & & -\frac{\tilde{s}_{12}}{\Delta\lambda_{31}} \\
 & & \frac{\tilde{c}_{12}}{\Delta\lambda_{32}} \\
\frac{\tilde{s}_{12}}{\Delta\lambda_{31}} & -\frac{\tilde{c}_{12}}{\Delta\lambda_{32}} & \\
\end{array}
\right).\label{eq:WP1}
\end{equation}

By the second order perturbation theory 
\begin{align}
(W^P_2)_{ij}=\begin{cases}
-\frac{1}{2}\sum\limits_{k\neq i}\frac{[2E(\check{H}_1)_{ik}]^2}{(\Delta\lambda_{ik})^2} &i=j \\
\frac{1}{\Delta\lambda_{ij}}\sum\limits_{k\neq i,k\neq j}\frac{2E(\check{H}_1)_{ik}2E(\check{H}_1)_{kj}}{\Delta\lambda_{kj}} &i\neq j
\end{cases},
\end{align}
then \\
\begin{widetext}
\begin{equation}
W^P_2=-\frac{(\epsilon^\prime \Delta m_{ee}^2)^2}{2} \left(\begin{array}{ccc}
\left(\frac{\tilde{s}_{12}}{\Delta\lambda_{31}}\right)^2 &-\frac{2\tilde{s}_{12}\tilde{c}_{12}}{\Delta\lambda_{32}\Delta\lambda_{21}} & 0 \\
\frac{2\tilde{s}_{12}\tilde{c}_{12}}{\Delta\lambda_{31}\Delta\lambda_{21}} & \left(\frac{\tilde{c}_{12}}{\Delta\lambda_{32}}\right)^2 & 0 \\
0 & 0 & \left(\frac{\tilde{s}_{12}}{\Delta\lambda_{31}}\right)^2+\left(\frac{\tilde{c}_{12}}{\Delta\lambda_{32}}\right)^2 \\
\end{array}
\right).\label{eq:WP2}
\end{equation}
\end{widetext}

First order corrections to the eigenvalues given by the perturbation theory is
\begin{equation}
\lambda^{P(1)}_i=2E(\check{H}_1)_{ii}=0,
\end{equation}
and second order corrections are
\begin{equation}
\lambda^{P(2)}_i=\sum\limits_{k\neq i}\frac{[2E(\check{H})_{ik}]^2}{\Delta\lambda_{ik}}.
\end{equation}
With Eq.~\ref{eq:Hbar} it is easy to get
\begin{align}
\lambda^{P(2)}_1&=-(\epsilon^\prime\Delta m^2_{ee})^2\frac{\tilde{s}^2_{12}}{\Delta\lambda_{31}}, \notag \\
\lambda^{P(2)}_2&=-(\epsilon^\prime\Delta m^2_{ee})^2\frac{\tilde{c}^2_{12}}{\Delta\lambda_{32}}, \notag \\
\lambda^{P(2)}_3&=(\epsilon^\prime\Delta m^2_{ee})^2\left(\frac{\tilde{s}^2_{12}}{\Delta\lambda_{31}}+\frac{\tilde{c}^2_{12}}{\Delta\lambda_{32}} \right).
\end{align}

\subsection{Perturbative expansion after the first two additional rotations}\label{Appendixasideexpansion}
After the first two additional rotations, we can implement a first order perturbative expansion to achieve second order accuracy for all eigenvalues and eigenstates\footnote{If we implement a perturbative expansion after only one additional rotation, it can be shown that one is required to do a second order expansion to achieve ${\cal O}(\epsilon^{\prime\,2})$ accuracy. Thus, starting the perturbative expansion one rotation earlier, as was done in \cite{Denton:2016wmg}, or performing an additional rotation before going to the perturbative expansion, as demonstrated in this Appendix, is more computationally efficient.}. 

For the eigenvalues this is evident. After the first two additional rotations, the eigenvalues $\lambda^{\prime\prime}_i$ ($\bar{\lambda}^{\prime\prime}_i$) already have the second order accuracy. Since diagonal entries of the perturbative Hamiltonian are always zero, a first order expansion will not give any corrections to the eigenvalues so the accuracy will be kept.

It is more complicated to test the eigenstates. In the following calculation we are assuming a case of neutrinos. We define 
\begin{equation}
(W^{P\prime\prime }_1)_{ij}\equiv -\frac{2E(\check{H}^{\prime\prime}_1)_{ij}}{\lambda^{\prime\prime}_{ij}},
\end{equation}
and all the diagonal elements of $W^{P\prime\prime}_1$ vanish. By Eq.~\ref{eq:checkH1pp} and Eq.~\ref{eq:alpha13} it can be figured out that
\begin{align}
W_1^{P\prime\prime}&=\epsilon^\prime\Delta m^2_{ee}\tilde{c}_{12}s_{\alpha_{13}}\begin{pmatrix}
& -\frac{c_{\alpha_{23}}}{\Delta\lambda^{\prime\prime}_{21}} & -\frac{s_{\alpha_{23}}}{\Delta\lambda^{\prime\prime}_{31}} \\
\frac{c_{\alpha_{23}}}{\Delta\lambda^{\prime\prime}_{21}} & & \\
\frac{s_{\alpha_{23}}}{\Delta\lambda^{\prime\prime}_{31}} & & \\
\end{pmatrix} \notag \\
\notag \\
&\simeq -(\epsilon^\prime\Delta m^2_{ee})^2\frac{\tilde{c}_{12}\tilde{s}_{12}}{\Delta\lambda_{31}}\begin{pmatrix}
& -\frac{1}{\Delta\lambda_{21}} &\quad \\
\frac{1}{\Delta\lambda_{21}}\!\!\! & & \\
& & \\
\end{pmatrix}+\mathcal{O}(\epsilon^{\prime\,3}).
\end{align}
Compared with Eq.~\ref{eq:threerotations}, we can get that
\begin{align}
&U_{13}(\alpha_{13})U_{23}(\alpha_{23})(\mathbb1+W^{P\prime\prime}_1) \notag \\
\simeq &\,U_{13}(\alpha_{13})U_{23}(\alpha_{23})U_{12}(\alpha_{12})+\mathcal{O}(\epsilon^{\prime\,3}).\label{eq:rotation_pert_aside}
\end{align}
So the eigenstates are corrected to second order accuracy.

For the case of anti-neutrinos, the perturbative Hamiltonian $\check{H}^{\prime\prime}_i$ will be different, so we need to re-calculate $W^{P\prime\prime}_1$ according to Table~\ref{tab:H_rotations}. Moreover, in Eq.~\ref{eq:rotation_pert_aside} $U_{13}(\alpha_{13})U_{23}(\alpha_{23})$ will be replaced by $U_{23}(\bar{\alpha}_{23})U_{13}(\bar{\alpha}_{13})$ and $\alpha_{12}$ will be replaced by $\bar{\alpha}_{12}$ .     

\section{Second order corrections to the mixing angles and CP phase} 
\label{Appendixsecondordercorrections}

The second order corrections to the mixing angles and CP phase, as defined in Eq.~\ref{eq:thetaexpansion} are
 \begin{align}
\tilde{\theta}^{(2)}_{13}={}&-\frac{\tilde{s}_{13}}{2\tilde{c}_{13}}\left[(W^\prime_1)_{23}\right]^2, \notag \\
\tilde{\theta}^{(2)}_{12}={}& (W^\prime_2)_{12}-\frac{\tilde{s}^2_{13}}{\tilde{c}^2_{13}}(W_1^\prime)_{13}(W_1^\prime)_{23},\notag \\
\tilde{\theta}^{(2)}_{23}={}&\frac{\tilde{c}_\delta \tilde{s}_{13}}{\tilde{c}^2_{13}}(W^\prime_1)_{13}(W^\prime_1)_{23}+\frac{c_{2\tilde{\theta}_{23}} \tilde{s}_\delta^2}{s_{2\tilde{\theta}_{23}}\tilde{c}^2_{13}}\left[(W^\prime_1)_{23}\right]^2,\notag \\
\tilde{\delta}^{(2)}={}&-\frac{2c_{2\tilde{\theta}_{23}}\tilde{s}_\delta \tilde{s}_{13}}{s_{2\tilde{\theta}_{23}}\tilde{c}^2_{13}}(W_1^\prime)_{13}(W_1^\prime)_{23}\notag\\
&+\frac{2(1+c_{2\tilde{\theta}_{23}}^2)\tilde{s}_\delta \tilde{c}_\delta}{\tilde{c}^2_{13} s_{2\tilde{\theta}_{23}}^2}[(W_1)_{23}^\prime]^2,
\end{align}
where in our case
\begin{align}
(W^\prime_1)_{13}={}& \epsilon^\prime\Delta m_{ee}^2 \tilde{s}_{12}\tilde{c}_{12} (\frac{1}{\Delta\lambda_{32}}-\frac{1}{\Delta\lambda_{31}}), \label{eq:W1P13}\\
(W^\prime_1)_{23}={}& \epsilon^\prime\Delta m_{ee}^2 (\frac{\tilde{s}^2_{12}}{\Delta\lambda_{31}}+\frac{\tilde{c}^2_{12}}{\Delta\lambda_{32}}), \label{eq:W1P23}
\end{align}
and
\begin{align}
(W^\prime_2)_{12}={}&(\epsilon^\prime\Delta m_{ee}^2)^2\tilde{s}_{12} \tilde{c}_{12}\left\{\frac{\tilde{c}^2_{12}}{\Delta\lambda_{32}\Delta\lambda_{21}}\right.\notag\\
&\left.+\frac{\tilde{s}^2_{12}}{\Delta\lambda_{31}\Delta\lambda_{21}}-\frac{1}{2}\left[\frac{\tilde{c}^2_{12}}{(\Delta\lambda_{32})^2}-\frac{\tilde{s}^2_{12}}{(\Delta\lambda_{31})^2}\right] \right\}. \label{eq:W2P12}
\end{align}
Actually $(W^\prime_1)_{ij}$ and $(W^\prime_2)_{ij}$ are elements of rotated $W_1$ and $W_2$ by $U_{12}(\tilde{\theta}_{12})$, i.e.
\begin{align}
W_1^\prime&\equiv U_{12}(\tilde{\theta}_{12})W_1U_{12}^\dagger(\tilde{\theta}_{12}), \notag \\
W_2^\prime&\equiv U_{12}(\tilde{\theta}_{12})W_2U_{12}^\dagger (\tilde{\theta}_{12}),
\end{align}
and they are invariants of a $\lambda_1\Leftrightarrow\lambda_2$ symmetry which will be explained in detail Appendix~\ref{Appendixsymmetry}.

Detailed formulas of the second order terms in Eq.~\ref{eq:sP} are

\begin{align}
f^{(2)}_{13}={}&-\frac{\tilde{s}_{13}}{2}\left[(W^\prime_1)^2_{13}+(W^\prime_1)^2_{23}\right] \notag \\
f^{(2)}_{12}={}&-\frac{\tilde{s}^2_{13}}{\tilde{c}^2_{13}}(W^\prime_1)_{23}\left[\frac{\tilde{s}_{12}}{2}(W^\prime_1)_{23}+\tilde{c}_{12}(W^\prime_1)_{13}\right]\notag\\
&+\tilde{c}_{12}(W^\prime_2)_{12} \notag \\
f^{(2)}_{23}={}&\frac{1}{\tilde{c}^2_{13}}(W^\prime_1)_{23}\left[\frac{c_{2\tilde{\theta}_{23}}\tilde{s}^2_{\delta}-\tilde{s}^2_{23}\tilde{c}^2_{\delta}}{2\tilde{s}_{23}}(W^\prime_1)_{23}\right.\notag\\
&\left.+\tilde{c}_{\delta}\tilde{s}_{13}\tilde{c}_{23}(W^\prime_1)_{13}\right] \notag \\
f^{(2)}_{\delta}={}&\frac{2\tilde{s}_{\delta}}{s_{2\tilde{\theta}_{23}}\tilde{c}^2_{13}}(W^\prime_1)_{23}\left[\frac{\tilde{c}^2_{\delta}(1+c^2_{2\tilde{\theta}_{23}})-\tilde{s}^2_{\delta}c^2_{2\tilde{\theta}_{23}}}{s_{2\tilde{\theta}_{23}}}(W^\prime_1)_{23}\right.\notag\\
&\left.-c_{2\tilde{\theta}_{23}}\tilde{c}_{\delta}\tilde{s}_{13}(W^\prime_1)_{13}\right]
\end{align}

The precision of the mixing angles through second order is shown in Fig.~\ref{fig:d1_angles}. It is evident that the approximated values achieve the expected accuracy.

\section{\texorpdfstring{$\lambda_1\Leftrightarrow\lambda_2$}{lambda1->lambda2} symmetry}\label{Appendixsymmetry}
If we exchange $\lambda_1$ and $\lambda_2$ and $\tilde{\theta}_{12}$ is translated to $\tilde{\theta}_{12}\pm\frac{\pi}{2}$, the Hamiltonian in basis of flavor eigenstates will keep unchanged because
\begin{multline}
\left(\begin{array}{cc}
\tilde{c}_{12}& \tilde{s}_{12} \\
-\tilde{s}_{12}& \tilde{c}_{12}
\end{array}
\right)\left(\begin{array}{cc}
\lambda_1& \\
&\lambda_2
\end{array}
\right)\left(\begin{array}{cc}
\tilde{c}_{12}& -\tilde{s}_{12} \\
\tilde{s}_{12}& \tilde{c}_{12}
\end{array}
\right)\\
=\left(\begin{array}{cc}
c_{(\tilde{\theta}_{12}\pm\frac{\pi}{2})}& s_{(\tilde{\theta}_{12}\pm\frac{\pi}{2})} \\
-s_{(\tilde{\theta}_{12}\pm\frac{\pi}{2})}& c_{(\tilde{\theta}_{12}\pm\frac{\pi}{2})}
\end{array}
\right)\left(\begin{array}{cc}
\lambda_2& \\
&\lambda_1
\end{array}
\right)\\
\times\left(\begin{array}{cc}
c_{(\tilde{\theta}_{12}\pm\frac{\pi}{2})}& -s_{(\tilde{\theta}_{12}\pm\frac{\pi}{2})} \\
s_{(\tilde{\theta}_{12}\pm\frac{\pi}{2})}& c_{(\tilde{\theta}_{12}\pm\frac{\pi}{2})}
\end{array}
\right).
\end{multline}
Under this discrete transformation
\begin{equation}
\lambda_1\Leftrightarrow\lambda_2,\quad \tilde{s}_{12}\Rightarrow-\tilde{c}_{12},\quad \tilde{c}_{12}\Rightarrow \tilde{s}_{12}.
\end{equation}
To the leading order
\begin{equation}
\alpha_{13}\Leftrightarrow\alpha_{23},
\end{equation}
which can be verified by Eq.~\ref{eq:alpha13} and Eq.~\ref{eq:alpha23}. $\tilde{\theta}_{23}$, $\tilde{\delta}$ and $\tilde{\theta}_{13}$ and their perturbing terms should be all invariants. Since it is a translation of $\tilde{\theta}_{12}$, the perturbation of $\tilde{\theta}_{12}$ should also be an invariant. Thus an implicit reason for introducing $W_1^\prime$ and $W_2^\prime$ can be revealed. It is easy to see in Eqs.~\ref{eq:W1P13}, \ref{eq:W1P23}, and \ref{eq:W2P12} that $W_1^\prime$ and $W_2^\prime$ are also invariants under the transformation. Then the perturbing terms are just combinations of some $\lambda_1\Leftrightarrow\lambda_2$ invariant functions.

\begin{figure}[t]
\centering
\includegraphics[width=\columnwidth]{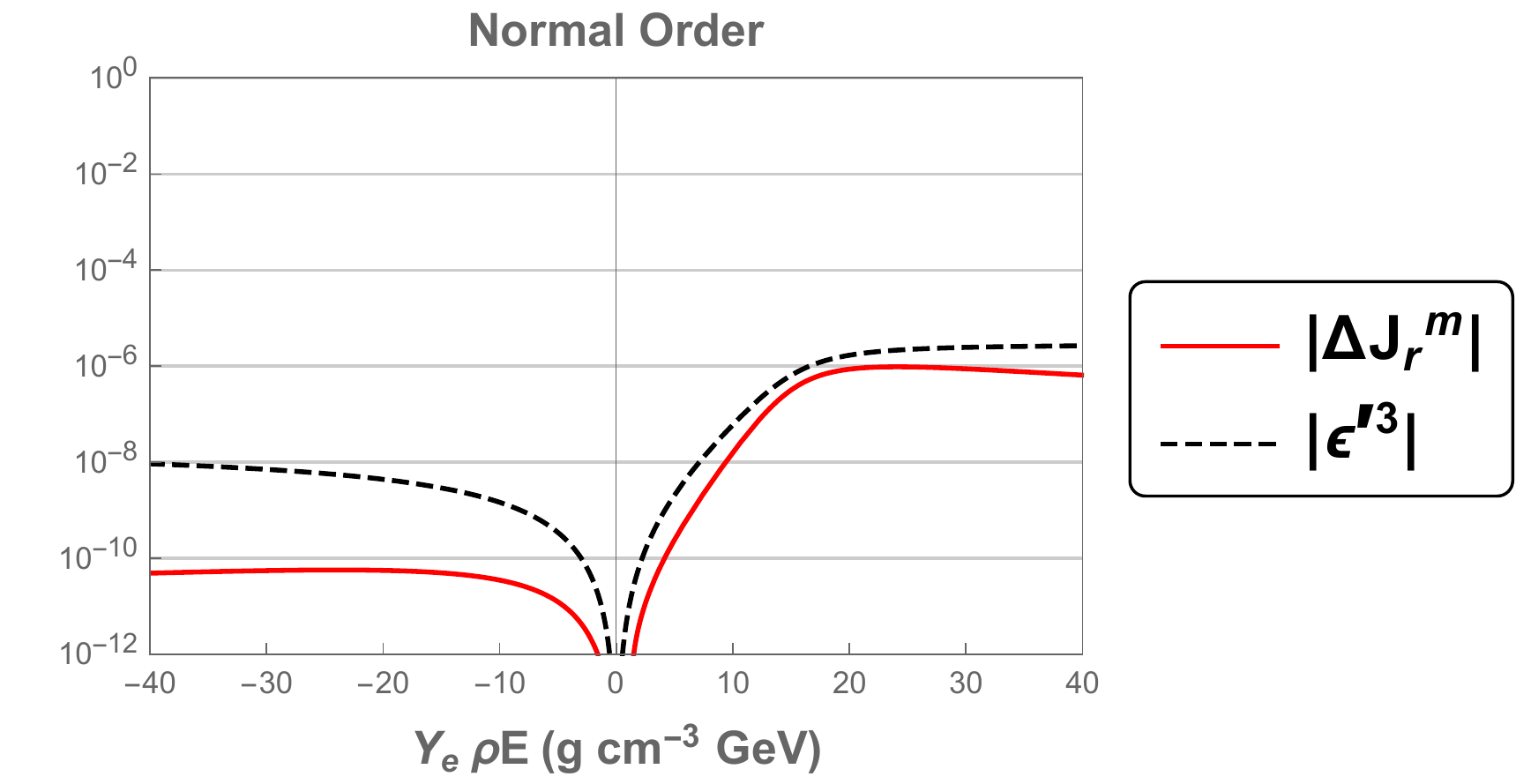}
\caption{This figure shows $\Delta J_r^m\equiv\tilde{s}^\prime_{12}\tilde{c}^\prime_{12}\tilde{s}^\prime_{13}\tilde{c}^{\prime 2}_{13}-s_{12}c_{12}s_{13}c^2_{13}\frac{\prod_{i>j}\Delta m^2_{ij}}{\prod_{i>j}\Delta\lambda^{\prime\prime\prime}_{ij}}$ through second order (red curve) for the normal mass ordering. The black dashed line is $\epsilon^{\prime3}$. }
\label{fig:deltaerror}
\end{figure}

\section{Some identities}\label{Appendixidentities}
In matter the corrected mixing angles, CP phase and eigenvalues must satisfy the Naumov-Harrison-Scott identity \cite{Naumov:1991ju,Harrison:1999df}, to second order, it is
\begin{multline}
s_{12}c_{12}s_{13}c^2_{13} s_{23}c_{23}s_{\delta}\prod_{i>j}\Delta m^2_{ij}\\
\simeq\tilde{s}^\prime_{12}\tilde{c}^\prime_{12}\tilde{s}^\prime_{13}\tilde{c}^{\prime 2}_{13}\tilde{s}^\prime_{23}\tilde{c}^\prime_{23}\tilde{s}^\prime_{\delta}\prod_{i>j}\Delta\lambda^{\prime\prime\prime}_{ij}+\mathcal{O}(\epsilon^{\prime 3}).
\end{multline}
A simpler identity is known as the Toshev identity \cite{Toshev:1991ku}, again to second order it is
\begin{equation}
s_{2\theta_{23}}s_{\delta}\simeq s^\prime_{2\tilde{\theta}_{23}}\tilde{s}^\prime_{\delta}+\mathcal{O}(\epsilon^{\prime 3}).
\label{eq:Toshev}
\end{equation}
Combining the above two identities a third identity can be derived \cite{Kimura:2002wd}
\begin{equation}
s_{12}c_{12}s_{13}c^2_{13}\frac{\prod_{i>j}\Delta m^2_{ij}}{\prod_{i>j}\Delta\lambda^{\prime\prime\prime}_{ij}}\simeq\tilde{s}^\prime_{12}\tilde{c}^\prime_{12}\tilde{s}^\prime_{13}\tilde{c}^{\prime 2}_{13}+\mathcal{O}(\epsilon^{\prime 3}).
\end{equation}
If we define 
\begin{align}
J_r&\equiv s_{12}c_{12}s_{13}c^2_{13}, \notag \\
J^m_r&\equiv \tilde{s}^\prime_{12}\tilde{c}^\prime_{12}\tilde{s}^\prime_{13}\tilde{c}^{\prime 2}_{13},
\end{align}
where $J_r$ is a reduced Jarlskog factor and similarly for the matter values, the third identity can be rewritten as
\begin{equation}
J_r\frac{\prod_{i>j}\Delta m^2_{ij}}{\prod_{i>j}\Delta\lambda^{\prime\prime\prime}_{ij}}\simeq J_r^m+\mathcal{O}(\epsilon^{\prime 3}).\label{eq:thirdidentity}
\end{equation}

For the third identity shown in Eq.~\ref{eq:thirdidentity}, analytical verification is complicated. An alternative numerical test is provided here. We define an error function as $\Delta J_r^m\equiv J_r^{m}-J_r\frac{\prod_{i>j}\Delta m^2_{ij}}{\prod_{i>j}\Delta\lambda^{\prime\prime\prime}_{ij}}$ to quantify the error in calculating the CP violating term using our expressions.
We have shown the precision of this expression in Fig.~\ref{fig:deltaerror}, in which we can see that the third identity holds to even better than third order in $\epsilon^\prime$.

\bibliography{Rotations}

\end{document}